\documentclass[twocolumn]{aastex631mod}
\usepackage{xspace}
\usepackage{amsmath}
\usepackage{enumerate}
\usepackage{comment}
\usepackage[nodayofweek]{datetime}
\newdateformat{monthyearday}{%
  \THEYEAR\ \monthname[\THEMONTH] \twodigit{\THEDAY}}

\makeatletter
\patchcmd{\NAT@citex}
  {\@citea\NAT@hyper@{%
      \NAT@nmfmt{\NAT@nm}%
      \hyper@natlinkbreak{\NAT@aysep\NAT@spacechar}{\@citeb\@extra@b@citeb}%
      \NAT@date}}
  {\@citea\NAT@nmfmt{\NAT@nm}%
    \NAT@aysep\NAT@spacechar\NAT@hyper@{\NAT@date}}{}{}

\patchcmd{\NAT@citex}
  {\@citea\NAT@hyper@{%
      \NAT@nmfmt{\NAT@nm}%
      \hyper@natlinkbreak{\NAT@spacechar\NAT@@open\if*#1*\else#1\NAT@spacechar\fi}%
        {\@citeb\@extra@b@citeb}%
      \NAT@date}}
  {\@citea\NAT@nmfmt{\NAT@nm}%
    \NAT@spacechar\NAT@@open\if*#1*\else#1\NAT@spacechar\fi\NAT@hyper@{\NAT@date}}
  {}{}
\makeatother

\newcommand{\rot}[1]{\rotatebox[origin=l]{90}{#1}}%
\newcommand{\sigmastar}{\ensuremath{\sigma_\star}}
\newcommand{\mstar}{\ensuremath{M_\star}}
\newcommand{\age}{\ensuremath{\mathrm{age}}\xspace}
\newcommand{\met}{\hbox{\ensuremath{\rm[Z/H]}}\xspace}
\newcommand{\afe}{\hbox{\ensuremath{\rm[\alpha/Fe]}}\xspace}
\newcommand{\med}[1]{\ensuremath{\left\langle\right.\!\!{#1}\!\!\left.\right\rangle}}
\newcommand{\EWOII}{\text{EW[O\:\textsc{ii}]}}
\newcommand{\OII}{\text{[O\:\textsc{ii}]}}
\newcommand{\OIII}{\text{[O\:\textsc{iii}]}}
\newcommand{\kms}{\ensuremath{\rm km\,s^{-1}}}
\defcitealias{Thomas2011}{TMJ11}

\newcommand*\ind[1]{%
  \ifcase#1\relax
    \ensuremath{\mathrm{Ca\:\textsc{ii}\:K}}
  \or 
    \ensuremath{\mathrm{Ca\:\textsc{ii}\:H}}
  \or 
    \ensuremath{\mathrm{D4000}}
  \or 
    \ensuremath{\mathrm{D_n4000}}
  \or 
    \ensuremath{\mathrm{H\delta_A}}
  \or 
    \ensuremath{\mathrm{H\delta_F}}
  \or 
    \ensuremath{\mathrm{CN_1}}
  \or 
    \ensuremath{\mathrm{CN_2}}
  \or 
    \ensuremath{\mathrm{Ca4227}}
  \or 
    \ensuremath{\mathrm{G4300}}
  \or 
    \ensuremath{\mathrm{H\gamma_A}}
  \or 
    \ensuremath{\mathrm{H\gamma_F}}
  \or 
    \ensuremath{\mathrm{Fe4383}}
  \or 
    \ensuremath{\mathrm{Ca4455}}
  \or 
    \ensuremath{\mathrm{Fe4531}}
  \or 
    \ensuremath{\mathrm{C_24668}}
  \or 
    \ensuremath{\mathrm{H\beta}}
  \or 
    \ensuremath{\mathrm{Fe5015}}
  \or 
    \ensuremath{\mathrm{Mg_1}}
  \or 
    \ensuremath{\mathrm{Mg_2}}
  \or 
    \ensuremath{\mathrm{Mg}\,b}
  \or 
    \ensuremath{\mathrm{Fe5270}}
  \or 
    \ensuremath{\mathrm{Fe5335}}
  \or 
    \ensuremath{\mathrm{Fe5406}}
  \or 
    \ensuremath{\mathrm{H/K\; ratio}}
  \or 
    \ensuremath{\mathrm{[MgFe]'}}
  \else
      \OtherCases 
  \fi
  \unskip\xspace}

\accepted{October 20, 2021}

\shorttitle{Passive Galaxies at Intermediate Redshift}
\shortauthors{Borghi et al. 2022a}
\graphicspath{{./}}

\begin{document}

\title{Toward a Better Understanding of Cosmic Chronometers: \\ Stellar Population Properties of Passive Galaxies at Intermediate Redshift}

\author[0000-0002-2889-8997]{Nicola Borghi}
\affiliation{Dipartimento di Fisica e Astronomia ``Augusto Righi''--Universit\`{a} di Bologna, via Piero Gobetti 93/2, I-40129 Bologna, Italy}
\affiliation{INAF---Osservatorio di Astrofisica e Scienza dello Spazio di Bologna, via Piero Gobetti 93/3, I-40129 Bologna, Italy}

\author[0000-0002-7616-7136]{Michele Moresco}
\affiliation{Dipartimento di Fisica e Astronomia ``Augusto Righi''--Universit\`{a} di Bologna, via Piero Gobetti 93/2, I-40129 Bologna, Italy}
\affiliation{INAF---Osservatorio di Astrofisica e Scienza dello Spazio di Bologna, via Piero Gobetti 93/3, I-40129 Bologna, Italy}

\author[0000-0002-4409-5633]{Andrea Cimatti}
\affiliation{Dipartimento di Fisica e Astronomia ``Augusto Righi''--Universit\`{a} di Bologna, via Piero Gobetti 93/2, I-40129 Bologna, Italy}
\affiliation{INAF---Osservatorio Astrofisico di Arcetri, Largo E. Fermi 5, I-50125, Firenze, Italy}

\author[0000-0003-4756-341X]{Alexandre Huchet}
\affiliation{Facult\'{e} des sciences d'Orsay---Universit\'{e} Paris-Saclay, 15 rue Georges Clemenceau, F-91405 Orsay cedex I, France}

\author[0000-0002-0449-8163]{Salvatore Quai}
\affiliation{Department of Physics and Astronomy, University of Victoria, 3800 Finnerty Road, Victoria, BC V8P 5C2, Canada}

\author[0000-0001-7085-0412]{Lucia Pozzetti}
\affiliation{INAF---Osservatorio di Astrofisica e Scienza dello Spazio di Bologna, via Piero Gobetti 93/3, I-40129 Bologna, Italy}

\correspondingauthor{Nicola Borghi}
\email{nicola.borghi6@unibo.it}

\begin{abstract}

We take advantage of the publicly available LEGA-C spectroscopic survey to measure the stellar population properties of 140 individual massive and passive galaxies at $z\sim0.7$. We develop and publicly release \texttt{PyLick}, a flexible Python code to measure UV to near-IR spectral indices. With \texttt{PyLick} we study the $\mathrm{H/K}$ ratio as a new diagnostic based on the pseudo-Lick Ca \textsc{II} H and K indices and find that a cut in $\mathrm{H/K}<1.1$ can be used jointly with other criteria to select (or verify the purity of) samples of passive galaxies. By combining photometric and spectroscopic criteria, we select a reliable sample of passively evolving galaxies. We constrain single-burst stellar ages, metallicities $\mathrm{[Z/H]}$, and $\mathrm{[\alpha/Fe]}$ with an optimized set of Lick indices, exploring in detail the robustness of our measurement against different combinations. Even without imposing cosmological priors, the derived ages follow a clear trend compatible with the expected cosmological aging of the Universe. We observe no significant redshift evolution for the metal abundance with respect to the values derived at $z=0$, with median $\mathrm{[Z/H]}=0.08\pm0.18$ and $\mathrm{[\alpha/Fe]}=0.13\pm0.11$. Finally, we analyze the relations between $\log\mathrm{age}$, $\mathrm{[Z/H]}$, $\mathrm{[\alpha/Fe]}$ and the stellar velocity dispersion, finding slopes of ($0.5\pm0.1$), ($0.3\pm0.2$), and ($0.2\pm0.1$), respectively; the small scatter of $<0.2$~dex points to rather homogeneous and short star formation histories. Overall, these results confirm and extend low-redshift findings of a mass-downsizing evolution. This work further strengthens the possibility of selecting pure samples of passive galaxies to be exploited reliably as cosmic chronometers to place independent cosmological constraints.

\end{abstract}

\keywords{Galaxy evolution (594), Galaxy abundances (574), Galaxy ages (576), Observational cosmology (1146)}


\section{Introduction}\label{introduction}
  The advent of deep spectroscopic surveys led to considerable progress in understanding galaxy formation and evolution over different cosmic epochs. The reason is that they enable us to combine two main observational techniques: look-back statistical studies of galaxy properties over different cosmic times, and the archaeological reconstruction of galaxy properties and star formation histories (SFHs) from spectroscopic data. 

  Galaxies with no or negligible levels of star formation (hereafter ``passive'' or ``quiescent'') are ideal laboratories to perform these studies because their stellar population is relatively simple to model \citep[for a detailed review, see][]{Renzini2006}. They are mostly found after the peak epoch of galaxy assembly (the so-called `cosmic noon'), especially below $z\sim 1.5$, when they dominate the stellar-mass density \citep{Muzzin2013, Ilbert2013}, and the number density of the most massive ones ($\mstar\gtrsim 10^{11}M_\odot$) remains almost constant between $z=1$ and $z = 0$ \citep{Pozzetti2010,Moresco2013}. On the contrary, at higher redshift, they constitute only a minor population \citep{Cimatti2004,Daddi2005}. Recent spectroscopic observations confirmed the existence of a few of these systems up to $z\sim 4$ \citep{Tanaka2019,Valentino2020,Santini2020} when the Universe was only $\sim 2$ Gyr old, requiring an early intense star formation followed by a complete decline (known as quenching). 

  Many different methods are used to investigate their physical properties. An example is the study of the photometric spectral energy distribution (SED; e.g., \citealt{Pacifici2016}). Photometric observations can span a very large range of wavelengths (usually from UV to FIR), allowing a broad view of various physical quantities, such as stellar mass, star formation rate, stellar age, and metallicity. However, it is not possible to find precise constraints due to strong degeneracies, e.g., age-metallicity \citep{Worthey1994}. To break them, high signal-to-noise spectroscopy is needed. The well-established method of Lick indices \citep{Faber1973, Burstein1984, Worthey1994} involves the use of a set of absorption features, each one having a unique sensitivity to stellar ages and element abundances. Another main approach is based on the use of the entire spectral information \citep[full spectral fitting; e.g.,][]{CidFernandes2005,Conroy2013a}. All of these techniques independently contributed to establishing the downsizing scenario \citep[first introduced by][]{Cowie1996}, according to which galaxy mass plays a major role in galaxy formation and evolution: massive galaxies are found to have evolved earlier and over shorter time scales than less massive ones. In the local Universe, this is supported by positive scaling relations between stellar ages, metallicities (\met), and $\alpha$-element abundances (\afe) with galaxy mass (dynamical and stellar) found in large, high-quality spectroscopic samples of early-type galaxies \citep{Kauffmann2003, Gallazzi2005,Gallazzi2006, Thomas2005,Thomas2010, Treu2005,Conroy2014, McDermid2015,Scott2017,Siudek2017,Gallazzi2021}. 
  
  The stellar metallicity, i.e. the amount of metals locked in stars, can shed light on the evolutionary stage of the stellar population, as well as on the external mechanisms that modify the chemical content of the interstellar medium. For instance, \cite{Peng2015} showed that local quiescent and star-forming galaxies form two distinct relations in the stellar-mass--metallicity plane. The difference between their shape can be interpreted either as an evidence for `strangulation'  \citep[i.e. the lack of new gas supply][]{Peng2015,Trussler2021}, or for shorter formation timescales coupled with galactic winds \citep{Spitoni2017} as the main mechanism driving galaxy quenching. The relative $\alpha$-element abundance with respect to iron is another key parameter. Core-collapsing massive stars are the main producers of $\alpha$-elements (O, Mg, Si, Ca, Ti), polluting the interstellar medium over relatively short timescales ($<100$~Myr). On the other hand, iron-peak elements (Fe, Cr) are primarily produced in Type Ia supernovae, which pollute the interstellar medium over longer timescales (100~Myr--2~Gyr). For this reason, the mean stellar \afe has been traditionally used as a SFH timescale diagnostic \citep{Thomas2005,Rosa2011}. In addition to stellar population properties, the environment can in principle have a role in shaping galaxy evolution. However, its effects seem weak once the correlation with mass is removed \citep{Thomas2010, LaBarbera2014, McDermid2015,Trussler2021}. To explore them in greater detail, very large samples of galaxies are needed \citep[e.g.,][]{Bluck2020}.

  At higher redshift, studies of stellar population properties are more challenging and require deep near-infrared spectroscopy. For this reason, they are mostly limited to samples of a few up to dozens galaxies \citep[e.g.,][]{Jorgensen2013,Gallazzi2014,Lonoce2015,Lonoce2020,Belli2019,Carnall2019, Kriek2019,Tacchella2021,Beverage2021}, and/or require the stacking of different galaxy spectra \citep[e.g.,][]{Choi2014,Onodera2015}. In particular, detailed studies of galaxy ages and chemical abundances for individual galaxies at intermediate redshift have been presented in very few works and usually by assuming the age of a $\Lambda$CDM universe as the maximum age allowed in the stellar population analysis. 
  
  Beyond galaxy evolution studies, massive and passive galaxies encode valuable information about the underlying cosmological framework. In fact, as first proposed by \cite{Jimenez2002}, it has been demonstrated that these objects can be used as \textit{cosmic chronometers} to trace the differential age evolution of the Universe $\mathrm{d}t/\mathrm{d}z$, and to provide in this way a cosmology-independent estimate of the expansion rate of the Universe $H(z)=-1/(1+z)\,\mathrm{d}z/\mathrm{d}t$. While we refer to the literature for a detailed discussion of the method and of the systematics involved \citep{Moresco2012b,Moresco2015,Moresco2016b,Moresco2020}, we underline here that there are two key ingredients to be met. First, the best cosmic chronometers that have been studied are very massive and passively evolving galaxies; therefore, it is fundamental to accurately select these objects by carefully excluding any possible hints of star-forming or young outliers. Second, it is crucial that in the differential age estimate $\mathrm{d}t$ no cosmological assumption is made. With the advantage of providing a cosmology-independent estimate of $H(z)$, the cosmic chronometers method has also been considered in several cosmological studies to place constraints on various cosmological models and parameters \citep{Moresco2012a,Seikel2012,Capozziello2014,Valkenburg2014,Sapone2014,Nunes2016,Sola2017,Moresco2017,Lhuillier2017,Yang2018,Haridasu2018,GomezValent2018,Lin2019,Lin2021}, with particular benefits over more standard cosmological probes \citep[see e.g.,][]{Vagnozzi2021}.

  In this work, we take advantage of the deep spectroscopic observations of the second data release of the Large Early Galaxy Astrophysics Census (LEGA-C DR2; \citealt{Wel2016, Straatman2018}) at $0.6\lesssim z \lesssim1$ to infer physical properties of individual passive galaxies without relying on any cosmological model or assumption. Our analysis is based on \texttt{Pylick}, a flexible Python tool to measure absorption features that we publicly release. It includes a wide set of indices already defined in the literature and also new diagnostics introduced and explored in this work for identifying passively evolving systems. Stellar population properties of each individual galaxy are derived with a Bayesian approach, adopting the simple stellar population (SSP) models of \cite{Thomas2011}. The analysis of their trends with redshift and stellar velocity dispersion will allow us to understand the individual and median properties of passive galaxies over different cosmic epochs and explore the underlying cosmology.

  The work is organized as follows. Section~\ref{sec:sample} gives an overview of the dataset, selection process, spectral index measurements, and main observational properties. Section~\ref{sec:analysis} presents background information on the models and the stellar population analysis. The main science results are presented and discussed in Section~\ref{sec:res}. A summary is presented in Section~\ref{sec:conclusions}. 

  In some cases, as reference values and mostly for illustrative purposes, we will use some theoretical relations based on cosmological models; for these cases, we adopt a `737 cosmology' (with $H_0=70$ km s$^{-1}$ Mpc$^{-1}$, $\Omega_m=0.3$, and $\Omega_\Lambda=0.7$).

\section{Passive galaxy sample}\label{sec:sample}
  \subsection{Data}
    The data used in this study are drawn from the second data release of LEGA-C, a recently completed ESO Public Spectroscopic Survey targeting $\sim$3000 $K_s$-band-selected galaxies at $0.6\lesssim z \lesssim 1$ in the COSMOS field. The observations were carried out with the Visible Multi-Object Spectrograph (VIMOS) on the VLT at Paranal Observatory. The flux-calibrated spectra span a wavelength range of $6300<\lambda <8800$~\AA\ with a spectral resolution of $R\sim3500$, and a median signal-to-noise ratio (S/N) of $\sim15$ per pixel (0.6~\AA). Spectra are obtained with $1\arcsec$ wide slits, corresponding to $\sim7$ kpc at these redshifts.

    A catalog of spectroscopic measurements for DR2 has been publicly released, comprising spectroscopic redshift, flux measurements for the main emission lines (Balmer lines, \OII$\lambda$3727, \OIII$\lambda$4959, \OIII$\lambda$5007), velocity dispersion determinations, as well as measurements for a sample of 14 Lick/IDS indices \citep{Straatman2018}. In our analysis, we consider the galaxy spectra, as well as the measurements of redshift ($z$), observed stellar velocity dispersion (\sigmastar), and \OII$\lambda$3727 emission-line flux. For the main analysis, we do not use the spectral indices measurements provided in the catalog, but instead determine our own line strengths from the spectra after matching the stellar population models resolution. This also allows us to extend the measurements to redder indices up to $\sim 5000$~\AA\ (see Section~\ref{sec:sample:indices}). The LEGA-C set of Lick indices will be used to validate our measurement code \texttt{PyLick} in \ref{app:pylick}.

    We cross-match the LEGA-C sample with the COSMOS2015 catalog \citep{Laigle2016} using a search radius of $1\arcsec$, to complement the spectroscopic information with photometric data, including bands NUV, $r$, and $J$, as well as stellar masses \mstar\ and (specific) star formation rates (${\rm sSFR=SFR}/\mstar$) derived through SED fitting. Finally, we use morphological information from the Zurich Estimator of Structural Types catalog \citep{Scarlata2007}, based on principal component analysis of the surface brightness profiles. Our parent sample is selected requiring good quality spectra (see \citealt{Straatman2018} for spectra quality flags) and available NUV, $r$, and $J$ absolute magnitudes. In this way, we end up with 1622 sources.

    \begin{figure*}
      \centering
      \includegraphics[width=.7\hsize]{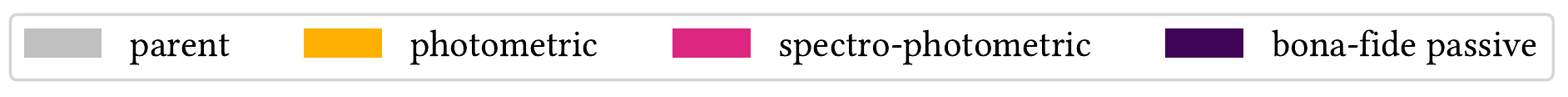}\\
      \includegraphics[width=.49\hsize]{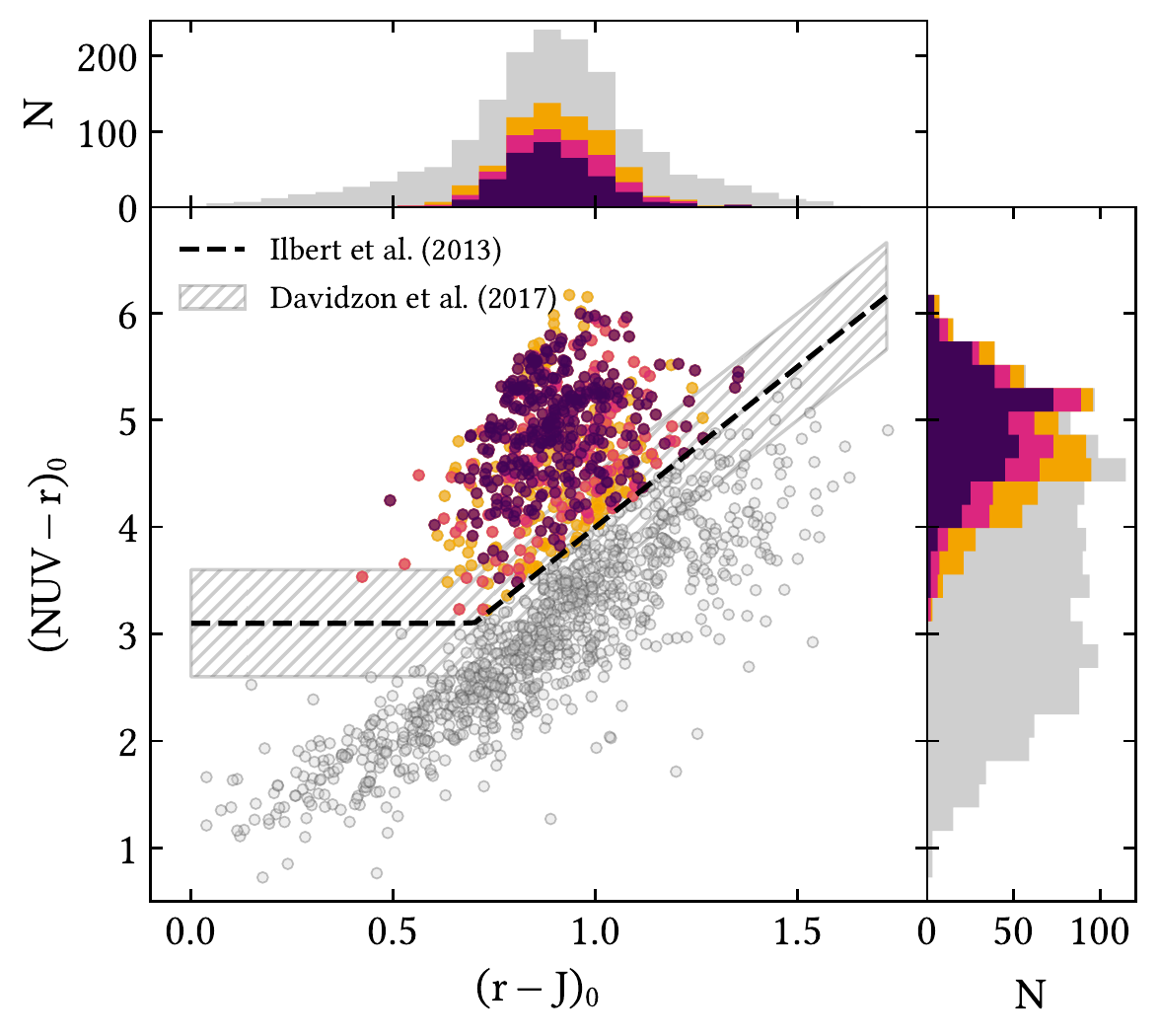}
      \includegraphics[width=.49\hsize]{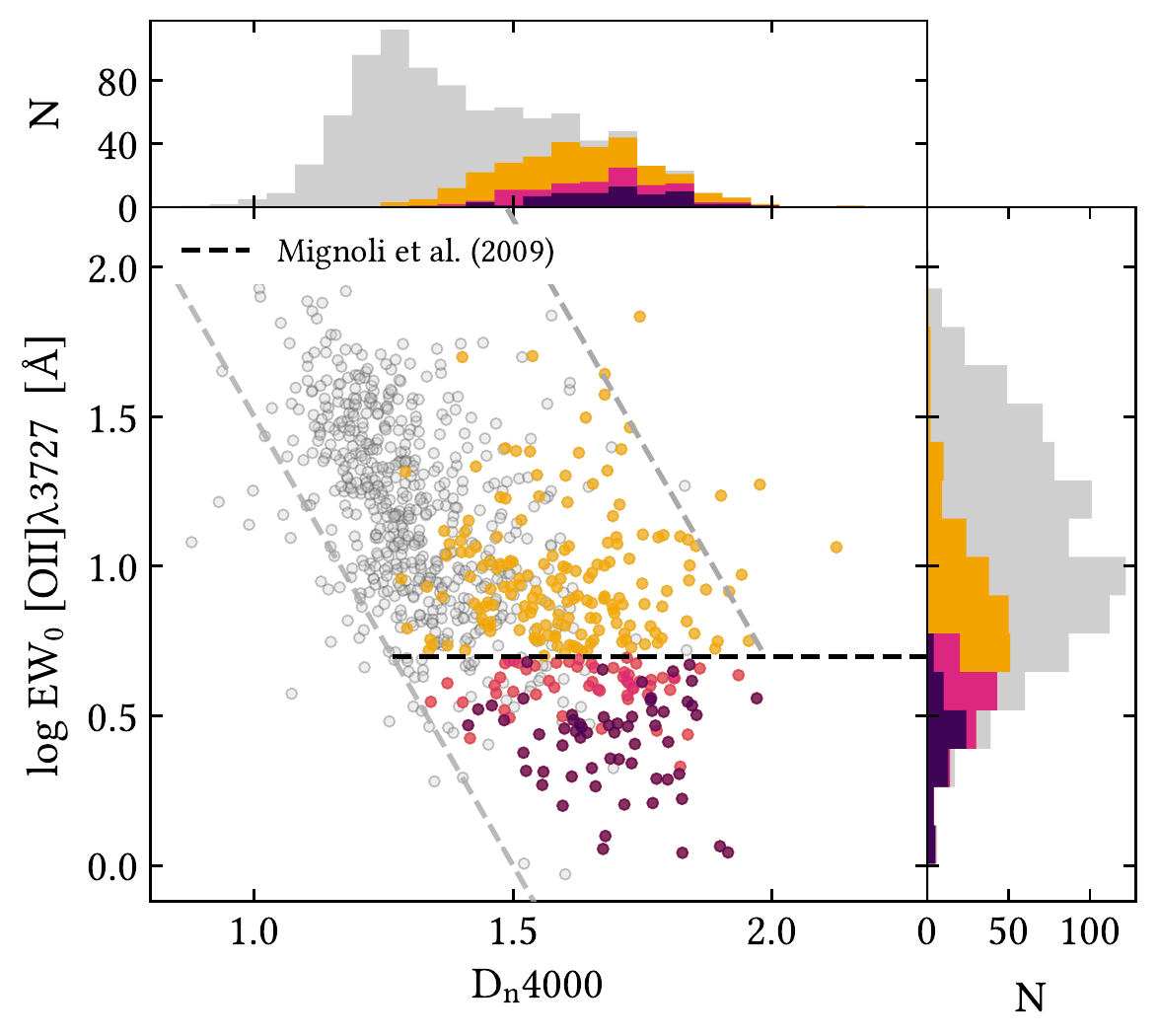}\\
      \includegraphics[width=\hsize]{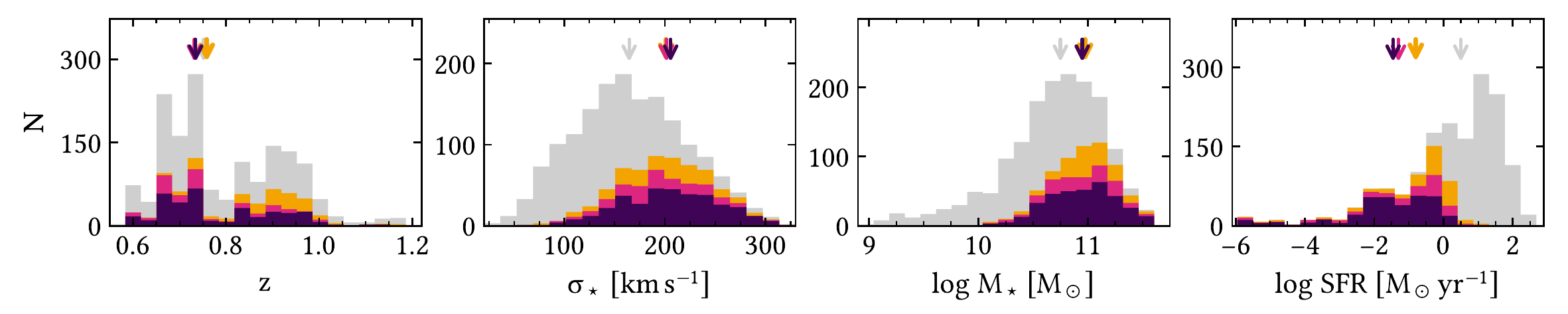}
      \caption{Distribution of the main properties of subsequently refined samples of passive galaxies in the LEGA-C survey. Upper panels: NUVrJ and \EWOII$-$\ind{3} diagnostic diagrams with selection criteria adopted in \citet{Ilbert2013} and \citet{Mignoli2009}, respectively. We use black colors for the criteria also adopted in this work, while gray lines are for illustrative purposes only. The hatched region ($\pm 0.05$~mag with respect to the NUVrJ cut) indicates the location of green-valley galaxies \citep{Davidzon2017}. Note that only 65/350 passive galaxies have detected \OII$\lambda$3727 line. Lower panels: distribution in redshift ($z$), observed stellar velocity dispersion ($\sigmastar$), stellar mass ($M_\star$), and star formation rate (SFR). Arrows represent the median values of the various subsamples.\label{fig:sample:selection_diagrams}}
    \end{figure*}

    \begin{deluxetable*}{cccccccccccc}
      \tablenum{1}
      \tablecaption{Median Properties of the Various Subsamples of Passive Galaxies Defined\label{table:sample_median}}
      \tablewidth{0pt}
      \tablehead{
        \colhead{Sample} & \colhead{$N$} & \colhead{\med{z}} & \colhead{\med{\sigmastar}} & \colhead{\med{\log\mstar}} &
        \colhead{\med{\log{\rm SFR}}} & \colhead{\med{\rm H/K}} & \colhead{\med{\ind{3}}} & \colhead{\med{\ind{4}}} &
        \colhead{\med{\ind{9}}} & \colhead{\med{\ind{12}}} & \colhead{\med{\ind{15}}} \\
        \colhead{} & \colhead{} & \colhead{} & \colhead{\kms} & \colhead{$M_\odot$} & \colhead{$ M_\odot\,\mathrm{yr^{-1}}$} & 
        \colhead{} & \colhead{} & \colhead{\AA} & \colhead{\AA} & \colhead{\AA} & \colhead{\AA} 
      }
      \colnumbers
      \startdata
      \textit{parent}                & 1622  & 0.752 & 165 & 10.75 &  0.50   & 1.166 & 1.426 & 2.941 & 2.357 & 2.583 & 3.256 \\
      \textit{photometric}           & 658   & 0.758 & 201 & 10.98 & -0.80  & 0.969 & 1.654 & 0.692 & 4.286 & 3.594 & 4.935 \\
      \textit{spectrophotometric}   & 485   & 0.732 & 202 & 10.95 & -1.30  & 0.957 & 1.681 & 0.462 & 4.387 & 3.741 & 4.996 \\
      \textit{bona fide passive}     & 350   & 0.735 & 206 & 10.95 & -1.46  & 0.957 & 1.690 & 0.347 & 4.443 & 3.832 & 5.097 \\
      \enddata
      \tablecomments{Data from: LEGA-C DR2 (3--4, \citealt{Straatman2018}); COSMOS2015 (5--6, \citealt{Laigle2016}); this work (7--12, Section~\ref{sec:sample:indices}).}
    \end{deluxetable*}
      
  \subsection{Selection criteria}\label{sec:sample:selection}
    In the literature, multiple methods have been proposed to separate `passive' from `star-forming' galaxies, including a morphological selection of spheroidal systems \citep[following the original separation by][]{Hubble1936}, cuts on color--color diagrams (e.g., UVJ, \citealt{Williams2009}; NUVrJ, \citealt{Ilbert2013}) or on a color--mass diagram \citep[e.g.,][]{Peng2010}, SED fitting \citep[e.g.,][]{Ilbert2009}, and sSFR criteria \citep[e.g.,][]{Pozzetti2010}.
    
    However, these different criteria do not perfectly overlap \citep{Renzini2006}. In fact, selections based on a single criterion are not stringent enough to reduce the contamination from star-forming outliers, with a percentage of contamination up to 10\%--30\% depending on the considered criterion. On the other hand, a combination of different criteria, maximizing the overlap of complementary information (photometric and spectroscopic), is significantly more effective in selecting a pure sample \citep{Franzetti2007,Moresco2013}. In this analysis, with the aim of studying the physical properties of cosmic chronometers, we are interested in having the purest possible sample, minimizing as much as possible the eventual residual contamination from star-forming outliers. We decided to combine different complementary cuts, as outlined below.

    \begin{itemize}
      \item\textbf{NUVrJ selection.} Photometric passive galaxies are selected using the rest-frame $\rm{NUV}-r)$ and $(r-J)$ colors following the criterion proposed by \citet{Ilbert2013}: $(\mathrm{NUV}-r)>3~(r-J)+1$ and $(\mathrm{NUV}-r)>3.1$. These colors have been demonstrated to be extremely sensitive to reveal objects with recent (1--100 Myr) star formation episodes (blue $\mathrm{NUV}-r$ colors) even if they are dust-obscured (redder $r-J$ colors), and are therefore optimized to safely separate quiescent and star-forming galaxies \citep{Arnouts2007, Ilbert2015, Davidzon2017}. With this cut, 658 sources are selected. We refer to these as the \textit{photometric} passive sample.
      \item\textbf{Emission-line cut.} We further restrict our sample by excluding galaxies with a significant \OII$\lambda$3727 emission line, a tracer of photoionized gas, which is typically considered an indicator of ongoing star formation\footnote{Low-ionization nuclear emission-line regions (LINERs) and ionization from old stars can also be responsible for \OII\ and \OIII\ emission-lines \citep[e.g.,][]{Yan2006,Singh2013,Cimatti2019Book}, but because we aim to sample purity we still exclude these sources.}. In particular, we exclude those galaxies that have an $\EWOII>5$~\AA. The threshold is found to separate well star-forming and passive galaxies in previous spectroscopic surveys at similar redshift \citep[e.g.,][]{Mignoli2009}. Combining this and the previous cut, we obtain 485 galaxies that we refer to as the \textit{spectrophotometric} passive sample.
      \item\textbf{Visual inspection.} The sample is further refined by visually inspecting all the remaining spectra. We remove galaxies with clearly strong \OII$\lambda$3727 and/or \OIII$\lambda$5007 lines, obtaining typical $\mathrm{S/N}<3$ in their EWs. The latter is crucial to spectroscopically characterize galaxies at $z\lesssim 0.65$ for which \OII\ is not available in LEGA-C spectra. Based upon these criteria, we obtain a final sample of 350 \textit{bona fide} passive galaxies.
    \end{itemize}
    
    In Figure~\ref{fig:sample:selection_diagrams} we show the distribution of LEGA-C galaxies in two diagnostic diagrams (NUVrJ, \citealt{Ilbert2013}; \EWOII$-$\ind{3} \citealt{Mignoli2009}), redshift, $\sigmastar$, $M_\star$, and SFR, across the \textit{parent}, \textit{photometric}, \textit{spectrophotometric} and \textit{bona fide} passive subsamples. The median values for the main quantities are quoted in Table~\ref{table:sample_median}. Note that objects without [OII] detection are not displayed in the upper-right diagram.
    
    Overall, the LEGA-C galaxies present two distinctive peaks in the redshift distribution at $z\sim0.7$ and $z\sim0.9$, with very few galaxies at $z>1$. They form two separate populations in the NUVrJ plot, with a blue sequence reaching low $(\mathrm{NUV}-r)\sim 1.1$ colors and a red cloud that constitutes the \textit{photometric} passive sample. Although a NUVrJ-only criterion drastically reduces the presence of star-forming systems in the sample ($\med{\rm \log sSFR/yr}=-11.8$), about one-third of objects still have a significant \OII\ emission. Therefore, a spectroscopic selection, here performed by combining a cut on the EW followed by a careful inspection of all the remaining \OII\ and/or \OIII\ contributions, is fundamental to ensure the purity of the sample. It is interesting to note that this does not only remove the tail of bluer $(\mathrm{NUV}-r)$ galaxies associated with the green-valley region, but also systems with redder colors. The final \textit{bona fide} passive sample has a median redshift of $\med{z}=0.735$. Passive galaxies are located toward the high-$\sigmastar$ and $\mstar$ tails of the parent distribution. In particular, the median  $\sigmastar$ ($\log \mstar/M_\odot$) increases from 164.5~\kms\ (10.75) to 205.7~\kms\ (10.95) moving from the \textit{parent} to the \textit{bona fide} passive sample, and most of the \textit{bona fide} passive galaxies ($85\%$) have $\log \mstar/M_\odot > 10.6$. With respect to the \textit{spectrophotometric}, this sample has a SFR lower by $0.16$ dex (with a median uncertainty of 0.18 dex). Finally, we note that the passive sample has a median specific star formation rate of $\med{\rm \log sSFR/yr}=-12.1$, with only 15 galaxies ($\sim4~\%$) reaching $>-11$, a value commonly adopted to classify ``passive'' galaxies \citep[see][]{Pozzetti2010}.

    \begin{figure*}[t!]
      \centering
        \includegraphics[width=.9\hsize]{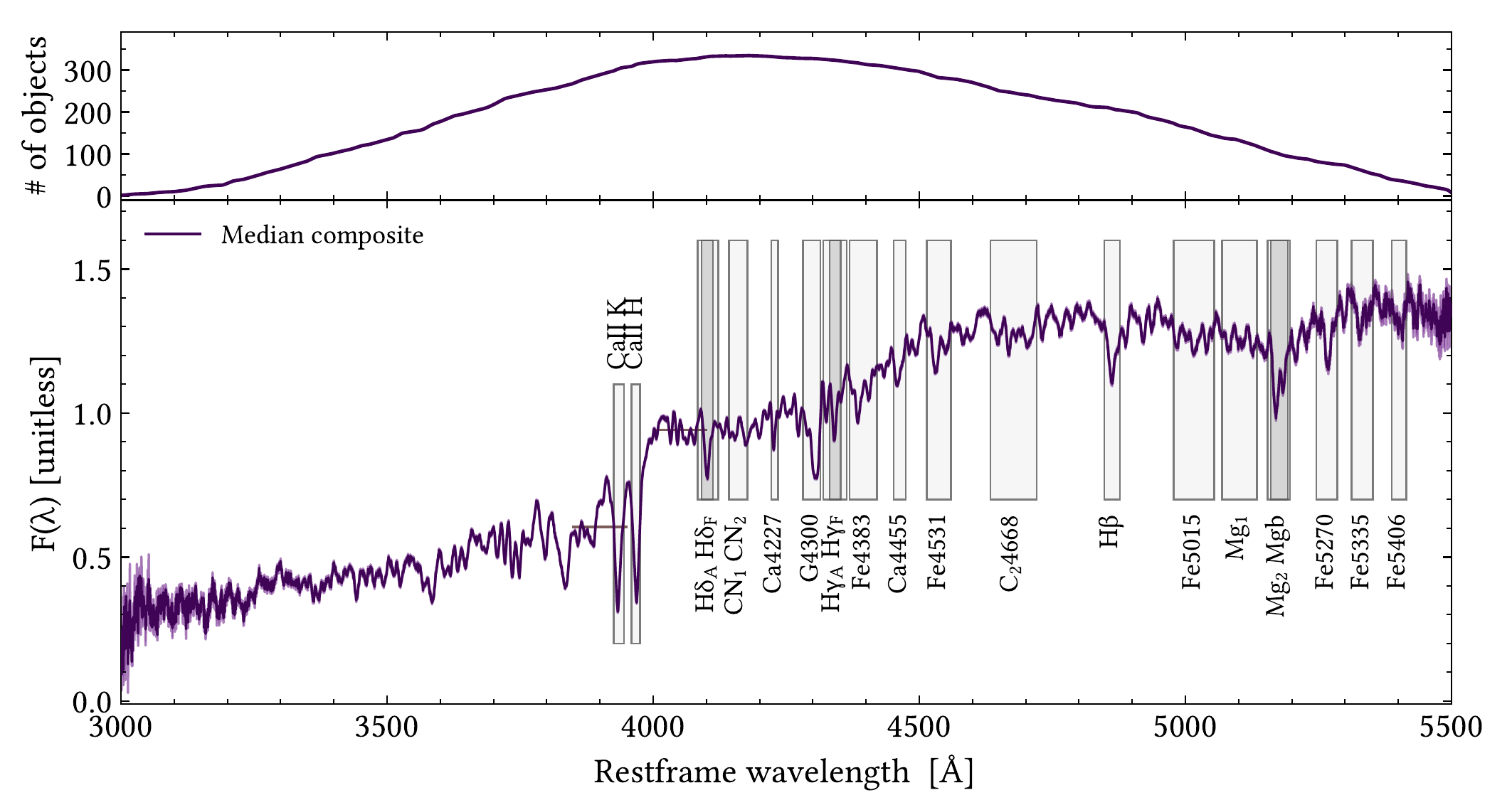}
        \caption{Median composite spectrum of the 350 selected passive galaxies. Upper panel: number of stacked galaxies at each wavelength. Lower panel: Median composite spectrum and associated 1$\sigma$ error. Gray boxes show central regions of the 22 Lick and pseudo-Lick indices measured in this work; narrower indices in overlapping regions (\ind{5}, \ind{11}, and \ind{20}) are shown with darker shading. Horizontal segments identify blue and red bandpasses where \ind{3} is computed.\label{fig:sample:composite_350_passive}}
    \end{figure*}

    We measure the composite spectrum of the \textit{bona fide} passive sample by normalizing each rest-frame spectrum to the median flux between 4200 and 4400~\AA, and interpolating it onto a common grid (3000--5500~\AA, $\Delta\lambda=0.35~\text{\AA}~\rm{pix^{-1}}$), considering only pixels with no problematic spectral flags. For each pixel, a $\sigma$-clipping is applied to reject fluxes $F_i$ deviating more than $4 \sigma$ from the mean, minimizing the impact of potential night sky emission-line residuals. The composite spectrum is obtained by computing the median flux and the associated uncertainty, given by the normalized median absolute deviation (NMAD\footnote{$\mathrm{NMAD}=1.4826\cdot\mathrm{median}(\left| F_i-\mathrm{median}(F_i)\right|)$, see \cite{Hoaglin1983}}) divided by the square root of the number of objects at each pixel. This technique provides a composite spectrum robust against imperfections and not biased toward galaxies with higher S/N. The resulting S/N per pixel measured around $4000$~\AA\ rest frame is $\sim 230$.

    In Figure~\ref{fig:sample:composite_350_passive} we show the median composite spectrum of the 350 passive galaxies selected in this work. The spectrum shows features typical of a passively evolving population. Qualitatively, we note a red continuum, indicative of an old stellar population; a significant 4000~\AA\ discontinuity; the absence of a Balmer break (3650~\AA); a \ind{0} absorption-line deeper than the associated \ind{1} (Section~\ref{sec:CaHK}); and the presence of several metallic indices (e.g., \ind{9}, \ind{12}, \ind{14}, and \ind{19}). It is very important to notice that even at the very high S/N of the stacked spectrum, no emission line is detectable, confirming the robustness of the selection. On the contrary, the composite spectrum of galaxies excluded by visual inspection has significant \OII\ and \OIII\ emission-lines.

    In conclusion, the distribution of SED-fitting-derived stellar masses and SFR, as well as the analysis of the composite spectrum, confirms the reliability of our selection of passive galaxies. These systems at $0.6<z<1$ show no detectable evidence of recent star formation. The presence of a possible underlying young component will be further assessed by studying spectral absorption features (Section~\ref{sec:sample:obs_prop}).

  \subsection{Measuring spectral indices with \texttt{PyLick}}\label{sec:sample:indices}
    In its complete version, the Lick system consists of 25 indices in the rest-frame wavelength interval $4000-6000$~\AA\ \citep{Worthey1997, Trager1998}. Each index requires the definition of a central region ($\lambda_{c1}$, $\lambda_{c2}$) and two other regions located toward the red ($\lambda_{r1}$, $\lambda_{r2}$) and blue ($\lambda_{b1}$, $\lambda_{b2}$) of the central one to estimate a reference pseudo-continuum level. Following the approach of the Lick group, the strenghts of atomic $I_a$ and molecular $I_m$ indices are calculated using the following equations: 
    \begin{align}
      &I_a(\text{[\AA]}) =\int_{\lambda_{c_1}}^{\lambda_{c_2}} \left( 1-\frac{F(\lambda)}{F_c(\lambda)}\right)\;\mathrm{d}{\lambda}, \label{eq:ind_licks1}  \\
      &I_m([\mathrm{mag}])=-2.5\log \left( \frac{1}{\lambda_{c_2}-\lambda_{c_1}}\int_{\lambda_{c_1}}^{\lambda_{c_2}}\frac{F(\lambda)}{F_c(\lambda)}\;\mathrm{d}{\lambda}\right), \label{eq:ind_licks2} 
    \end{align}
    where $F(\lambda)$ and $F_c(\lambda)$ are the spectrum flux and the local pseudo-continuum, respectively. The latter is commonly derived through linear interpolation:
    \begin{equation}\label{eq:3lickpseudocont}
      F_c(\lambda) = F_b \frac{\lambda_r -\lambda}{\lambda_r -\lambda_b} + F_r \frac{\lambda -\lambda_b}{\lambda_r -\lambda_b},
    \end{equation}
    where $\lambda_{b,r}$ and $F_{b,r}$ are the central wavelengths and mean fluxes of the lateral blue and red regions.

    \begin{deluxetable}{lccccc}
      \tablenum{2}
      \tablecaption{Spectral Indices Properties for the 350 Passive Galaxies\label{tab:ind_prop}}
      \tablewidth{0pt}
      \tablehead{
        \colhead{Index} & 
        \colhead{$z^{(a)}$} & 
        \colhead{Range$^{(b)}$} & 
        \colhead{$\med{I}^{(c)}$} &
        \colhead{\med{S/N}$^{(c)}$} 
        }
      \startdata
      \ind{0}  & $>0.65$  &  { }4.928--9.157  & { }7.167  & 22.40 \\
      \ind{1}  & $>0.65$  &  { }5.472--8.340  & { }6.891  & 40.93 \\
      \ind{2}  & $>0.65$  &  { }1.732--2.028  & { }1.881  & 236.61 \\
      \ind{3}  & $>0.65$  &  { }1.517--1.846  & { }1.690  & 172.13 \\
      \ind{4}  & All      &   --1.438--3.826  & { }0.347  & 3.44   \\
      \ind{5}  & All      &  { }0.630--3.516  & { }1.709  & 10.37  \\
      \ind{6}  & All      &   --0.039--0.098  & { }0.035  & 5.34   \\
      \ind{7}  & All      &   --0.005--0.145  & { }0.074  & 9.21   \\
      \ind{8}  & All      &  { }0.290--1.701  & { }0.922  & 7.69   \\
      \ind{9}  & All      &  { }2.605--5.818  & { }4.443  & 18.29  \\
      \ind{10} & All      &   --4.225--0.431  &  --2.729  & 9.86   \\
      \ind{11} & All      &   --1.121--2.348  &  --0.105  & 3.64   \\
      \ind{12} & All      &  { }1.675--5.778  & { }3.832  & 11.87  \\
      \ind{13} & $<0.96$  &  { }0.219--2.366  & { }1.395  & 8.54   \\
      \ind{14} & $<0.92$  &  { }1.438--4.599  & { }2.958  & 11.16  \\
      \ind{15} & $<0.88$  &  { }1.423--7.925  & { }5.097  & 13.94  \\
      \ind{16} & $<0.77$  &  { }1.175--3.490  & { }1.989  & 13.72  \\
      \ind{17} & $<0.74$  &  { }1.641--7.171  & { }4.737  & 13.98  \\
      \ind{18} & $<0.68$  &  { }0.036--0.145  & { }0.077  & 23.94  \\
      \ind{19} & $<0.68$  &  { }0.079--0.280  & { }0.205  & 46.88  \\
      \ind{20} & $<0.69$  &  { }1.789--4.827  & { }3.274  & 17.08  \\
      \ind{21} & $<0.64$  &  { }0.628--4.054  & { }2.549  & 15.96  \\
      \ind{22} & $<0.63$  &  { }0.425--3.553  & { }2.448  & 13.02  \\
      \ind{23} & $<0.62$  &  { }0.144--2.462  & { }1.573  & 8.67   \\
      \enddata
      \tablecomments{$(a)$ Expected redshift coverage within $0.6<z<1$ of the VIMOS HR red spectrograph; $(b)$ Computed between $5^{th}-95^{th}$ percentiles; $(c)$ Median index value and signal-to-noise ratio. Index units are angstr\"{o}m for all indices except: \ind{6}, \ind{7}, \ind{18}, \ind{19} (mag); \ind{2}, \ind{3} (dex).}
      \vspace*{-3em}
    \end{deluxetable}

    \vspace{-1em}
    One of the strongest features in passive galaxy spectra is the 4000~\AA\ discontinuity (D4000). At bluer wavelengths, the flux suddenly declines due to the accumulation of a large number of spectral lines that are present in stellar types cooler than G0. This property makes it a good age tracer \citep[e.g.,][]{Kauffmann2003,Moresco2011}. A discontinuity index can be quantified as the ratio of the average flux density in two regions redwards and bluewards of 4000~\AA:
    \begin{equation}\label{eq:ind_D4000} 
      \ind{2}=\frac{\lambda_{b2}-\lambda_{b1}}{\lambda_{r2}-\lambda_{r1}} \frac{ \int_{\lambda_{r1}}^{\lambda_{r2}}\lambda^2 F(\lambda) \;\mathrm{d}{\lambda}} { \int_{\lambda_{b1}}^{\lambda_{b2}} \lambda^2 F(\lambda)\;\mathrm{d}{\lambda}}
    \end{equation}
    where $\lambda_{b1}, \lambda_{b2} = 3750, 3950$~\AA\ and $\lambda_{r1}, \lambda_{r2} = 4050, 4250$~\AA\ in the original definition by \cite{Bruzual1983}. Alternatively, \cite{Balogh1999} introduced a narrower definition of the \ind{2} (\ind{3}, estimated over the ranges 3850--3950 and 4000--4100~\AA), which is optimized to be less sensitive to reddening effects.

    To perform all the measurements of absorption features, we developed a dedicated Python code named \texttt{PyLick}. It is a public\footnote{The code is available at \url{https://gitlab.com/mmoresco/pylick/}} flexible Python library developed for the fast and accurate estimation of the main spectral indices introduced in the literature from the UV to the near-IR. A more detailed description of the main elements of the code, as well as validation tests against available LEGA-C DR2 data, can be found in \ref{app:pylick}. In this work, we will focus on atomic and molecular (Lick) indices and \ind{2}. Because spectral indices are defined on fixed wavelength intervals, the measurement can be affected by systematics if spectral broadening effects (e.g., instrumental resolution, stellar velocity dispersion) are not properly taken into account. In general, this results in lower index values compared to intrinsic ones. To address this issue, we convolve each rest-frame spectrum ($\mathrm{FWHM}_\mathit{spec}\approx 1.3$~\AA\ at $z=0.7$) to match the resolution of the models that will be implemented ($\mathrm{FWHM}_\mathit{mod}=2.5$~\AA), by using a Gaussian kernel with a standard deviation of $\sigma=(\mathrm{FWHM}_\mathit{model}^2-\mathrm{FWHM}_\mathit{spec}^2)^{1/2}/2.355$. After the measurement, we calibrate indices to zero-velocity dispersion following the approach described by \citet{Carson2010}. The procedure to derive the correction coefficients is described in detail in Huchet et al. (2022, \textit{in preparation}). Briefly, we measure indices on SSP spectra broadened at different $\sigma_\star$ generated from the 2016 version of \citet{Bruzual2003} models using the MILES stellar spectral library \citep{FalconBarroso2011} and \citet{Chabrier2003} initial mass function\footnote{The models are available at \url{http://www.bruzual.org/bc03/Updated_version_2016/}} and fit each index--$\sigma_\star$ relation with a fourth-order polynomial. 
    
    At the end of this process, we obtain a set of 24 spectral indices, namely:
    \begin{enumerate}\itemsep1pt
      \item[-] Balmer indices: \ind{4}, \ind{5}, \ind{10}, \ind{11}, \ind{16};
      \item[-] Iron-dominated indices: \ind{12}, \ind{14}, \ind{17}, \ind{21}, \ind{22};
      \item[-] Molecular indices: \ind{6}, \ind{7}, \ind{18}, \ind{19};
      \item[-] Other Lick indices available in the LEGA-C spectral range: \ind{8}, \ind{9}, \ind{14}, \ind{15}, \ind{20}; 
      \item[-] 4000~\AA\ discontinuity indices: \ind{2}, \ind{3};
      \item[-] Two recently defined pseudo-Lick indices: \ind{0}, \ind{1} (Section~\ref{sec:CaHK}).   
    \end{enumerate}

    \begin{deluxetable}{cccc}
      \tablenum{3}
      \tablecaption{Definitions of Pseudo-Lick \ind{0} and \ind{1} Indices Introduced in \citet{Fanfani2019}\label{tab:HKdef}}
      \tablewidth{0pt}
      \tablehead{
        \colhead{Index} & 
        \colhead{Central} & 
        \colhead{Blue} & 
        \colhead{Red} \\
        \colhead{} & 
        \colhead{\AA} & 
        \colhead{\AA} & 
        \colhead{\AA}      
        }
      \startdata
      \ind{0} & 3925.65 -- 3945 & 3845 -- 3880 & 3950 -- 3954 \\
      \ind{1} & 3959.40 -- 3978 & 3950 -- 3954 & 3983 -- 3993 \\
      \enddata
    \end{deluxetable}

    \newpage
    \vspace*{-3em}  
    This dataset extends LEGA-C DR2 public catalog from \ind{16} to \ind{23} indices and, in particular, Mg ones, commonly used as proxies to study the $\alpha$ enhancement. For each index, we compute redshift coverage, 5th--95th percentile range, median value, and median S/N. These values are presented in Table~\ref{tab:ind_prop}. 
    
    Trivially, bluer (redder) indices are available only at higher (lower) redshift. This information, combined with the upper panel of Figure~\ref{fig:sample:composite_350_passive}, gives an idea of the spectral and redshift coverage of this study. Most of the indices are available only for a narrow rest-frame wavelength range between 3900--4500~\AA. However, a significant number of galaxies ($\sim 200$) still share a common wavelength range between 3700--4900~\AA. In this case, the redshift coverage is reduced to $z\lesssim 0.9$. The median S/N of the measured indices is $> 10$, except for those with a signal $\sim 0$ (Balmer and CN indices), or those defined on a narrow central region (\ind{8}, \ind{13})

    \begin{figure}
      \centering
      \includegraphics[width=\hsize]{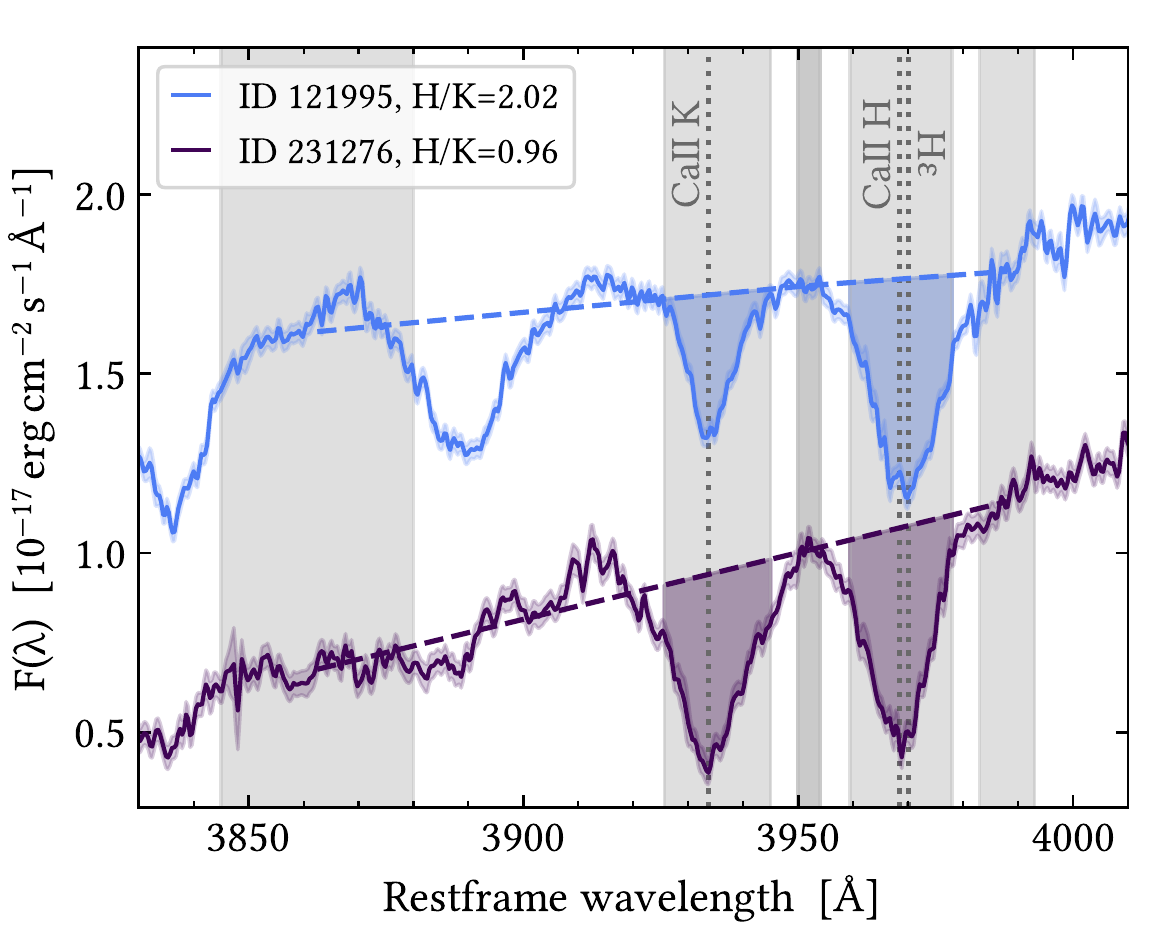}
          \caption{Measurement of pseudo-Lick indices \ind{0} and \ind{1} of a \textit{bona fide} passive galaxy (violet) and a NUVrJ star-forming galaxy (blue, with a vertical offset of $+0.35$). Vertical bands show the regions where these indices are defined (red pseudo-continuum of \ind{1} and blue pseudo-continuum of \ind{0} are overlapped). Vertical dotted lines are the centroids of \ind{0}, \ind{1}, and H$\epsilon$ lines. Dashed segments show the estimated pseudo-continuum flux used as a reference to compute the indices values (filled regions).\label{fig:sample:HK_definition}}
    \end{figure}

  \subsubsection{The H/K ratio}\label{sec:CaHK}
    \cite{Rose1984} firstly proposed to study stellar populations using the ratio of the minimum fluxes of two nearby lines. Specifically, \ind{1} over \ind{0} ratio was found to correlate with the starburst ages in post-starburst galaxies. As a matter of fact, H$\epsilon$ absorption line ($\lambda=3970.1$~\AA), which is deeper in the presence of young A- and B-type stars, overlaps to \ind{1} ($\lambda=3968.5$~\AA), while \ind{0} ($\lambda=3933.7$~\AA) remains relatively uncontaminated. The nomenclature H/K is thus a compact notation for $(\ind{1}+{\rm H}\epsilon)/\ind{0}$ and can be used as a stellar population diagnostic. Even the presence of a fraction of a young ($\lesssim 1$~Gyr) stellar component can significantly alter this spectral feature \citep{Longhetti1999, Lonoce2014, Moresco2018}. In the literature, this value has been usually measured as the ratio of minimum fluxes in H and K lines \citep[e.g.,][]{Rose1985,Leonardi1996,Longhetti1999, Lonoce2014, Moresco2018}:
    \begin{equation}\label{eq:HKmin}
      \left| \mathrm{H/K}\right|_\mathrm{min} = \frac{F_\mathrm{min}({\rm H})}{F_\mathrm{min}({\rm K})} .
    \end{equation}
    This method does not rely on measuring integrated quantities (differently from Lick indices), hence it is relatively independent of changes in spectral broadening, like those due to the instrumental resolution and stellar velocity dispersion. However, measurements of minima can be strongly biased by the presence of noise peaks, especially in low-S/N regimes. For this reason, we adopt here a hybrid approach recently introduced by \cite{Fanfani2019}, where H/K is computed as a ratio of two pseudo-Lick indices,
    \begin{equation}\label{eq:HKew}
      \mathrm{H/K} = \frac{ I_{\rm H}}{I_{\rm K}},
    \end{equation}
    using the \ind{0} and \ind{1} index passbands listed in Table~\ref{tab:HKdef} that have also been included in \texttt{PyLick}. We measure and \sigmastar-correct these indices to derive H/K values and associated uncertainties for the entire \textit{parent} sample. Two representative examples are shown in Figure~\ref{fig:sample:HK_definition}. For a typical passive population, the H line is less deep than the K line, therefore $\left| \mathrm{H/K}\right|_\mathrm{min}>1$. The so-called H/K inversion can already arise for a contribution of $\sim 5$~\% in mass of a young stellar population (with ages $<200$~Myr; \citealt{Moresco2018}). The two quantities defined in Eqs.~\ref{eq:HKmin} and \ref{eq:HKew} have an inverse relationship, but there is not a strict one-to-one correspondence. From the parent sample, we find an equivalent inversion value for the $\mathrm{H/K}$ between 1.2 and 1.5; therefore, the passive regime corresponding to $\left| \mathrm{H/K}\right|_\mathrm{min}> 1$ can be safely defined at $\mathrm{H/K} < 1.2$.

    \begin{figure*}[t]
      \centering
        \includegraphics[width=0.98\hsize]{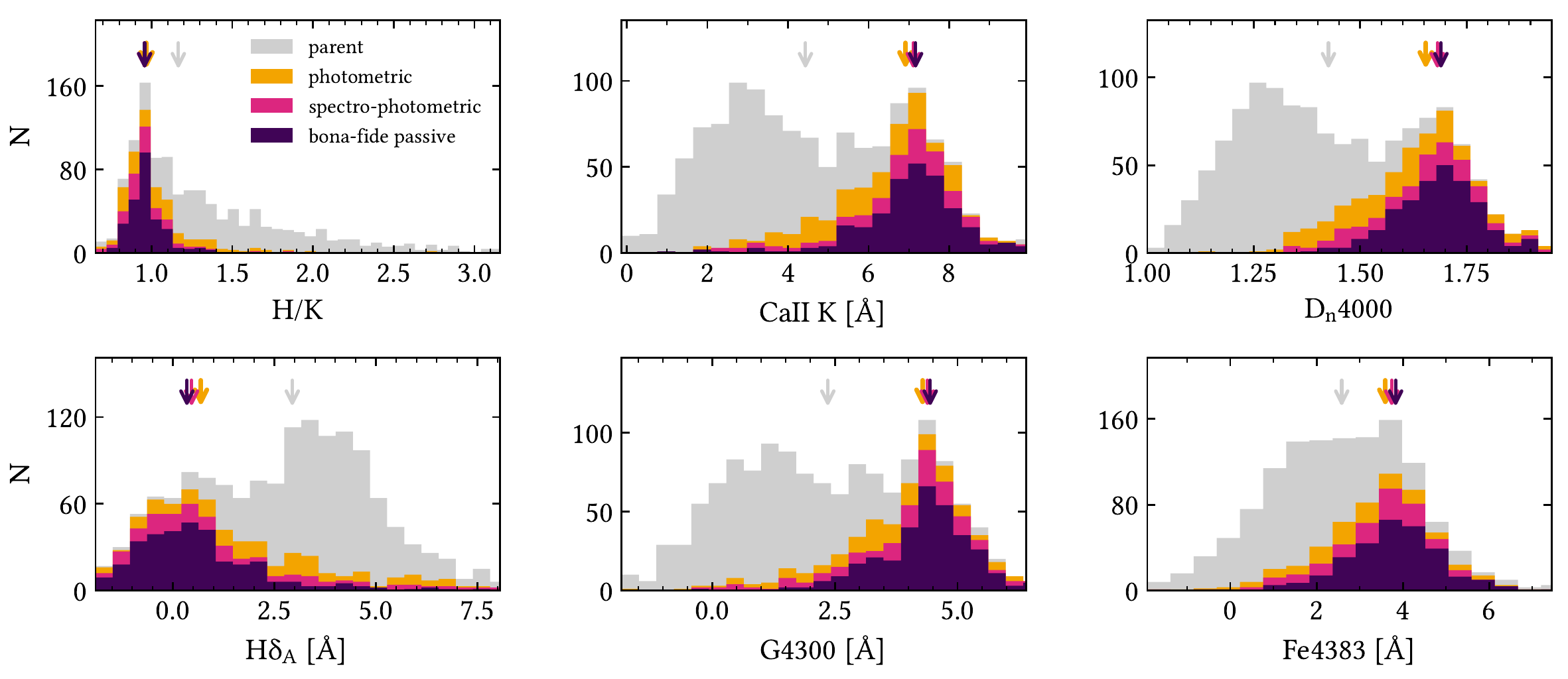}
      \caption{Distribution of LEGA-C galaxies as a function of the main spectral indices measured in this work. Different colors are used for the \textit{parent} (gray),  \textit{photometric} (yellow), \textit{spectrophotometric} (pink) and \textit{bona fide passive} (violet) subsamples. Arrows represent the median values.\label{fig:sample:obs_prop_3x2}} 
    \end{figure*}

  \subsection{Observed spectral features}\label{sec:sample:obs_prop}
    The distributions of the main measured absorption features for the various subsamples of passive galaxies are presented in Figure~\ref{fig:sample:obs_prop_3x2}, and the corresponding median values are quoted in Table~\ref{table:sample_median}. 

    Overall, the \textit{bona fide} passive galaxies tend to segregate with respect to the \textit{parent} sample median. We observe typical $\rm{H/K}$ ratios of $0.96\pm 0.08$ and very few systems with $\rm{H/K}>1.1$, while parent sample galaxies reach values well above $1.2$. It is important to stress here that even without applying any cut on the $\rm{H/K}$ ratio, our selection criteria allows us to discard the majority of the tail at high H/K, leaving us a sample with values characteristic of a passive population. This is important and independent evidence supporting the purity of our selection and the negligible contamination from underlying young stellar populations in the passive sample. Another interesting result that we report for the first time is that the \ind{0} line itself shows a clear bimodality, with passive galaxies having $\ind{0}>5$~\AA. This is not the case for \ind{1}, because as mentioned above, the H$\epsilon$ line strengthens the index in younger populations. We note here that \ind{0} is likely to be one of the main drivers of the D4000 bimodality as it is included in the blue passband of the D4000 index, while the \ind{1} shows a much flatter trend as a function of redshift. The well-known \ind{3} and H$\delta$ bimodalities are already studied in detail in the local universe \citep{Kauffmann2003, Siudek2017} and in LEGA-C data \citep{Wu2018a}. Photometrically selected massive and passive galaxies already populate high-\ind{3} and low-\ind{4} tails, indicating relatively old stellar populations. The addition of a spectroscopic criterion removes a significant number of remaining low-\ind{3} and high-\ind{4} galaxies. Among \textit{bona fide} passive galaxies, only 11 ($3\%$) have $\ind{3}<1.5$, and only 28 ($8\%$) have  $\ind{4}>2.5$~\AA. However, we underscore here that no additional cuts have been applied to minimize selection biases. Interestingly, we observe a slight bimodality also for \ind{9}, which measures the optical CH band (also known as G-band) and is very sensitive to the carbon abundance \citep{Tripicco1995,Korn2005}. The \textit{bona fide} passive galaxies have a relatively high $\ind{9}\gtrsim 2.5$~\AA. Finally, they are characterized by high \ind{12}, a primary indicator of stellar metallicity, with typical values of $\med{\ind{12}}\sim 3.8$~\AA.

    \begin{figure*}
      \centering
        \includegraphics[width=\hsize]{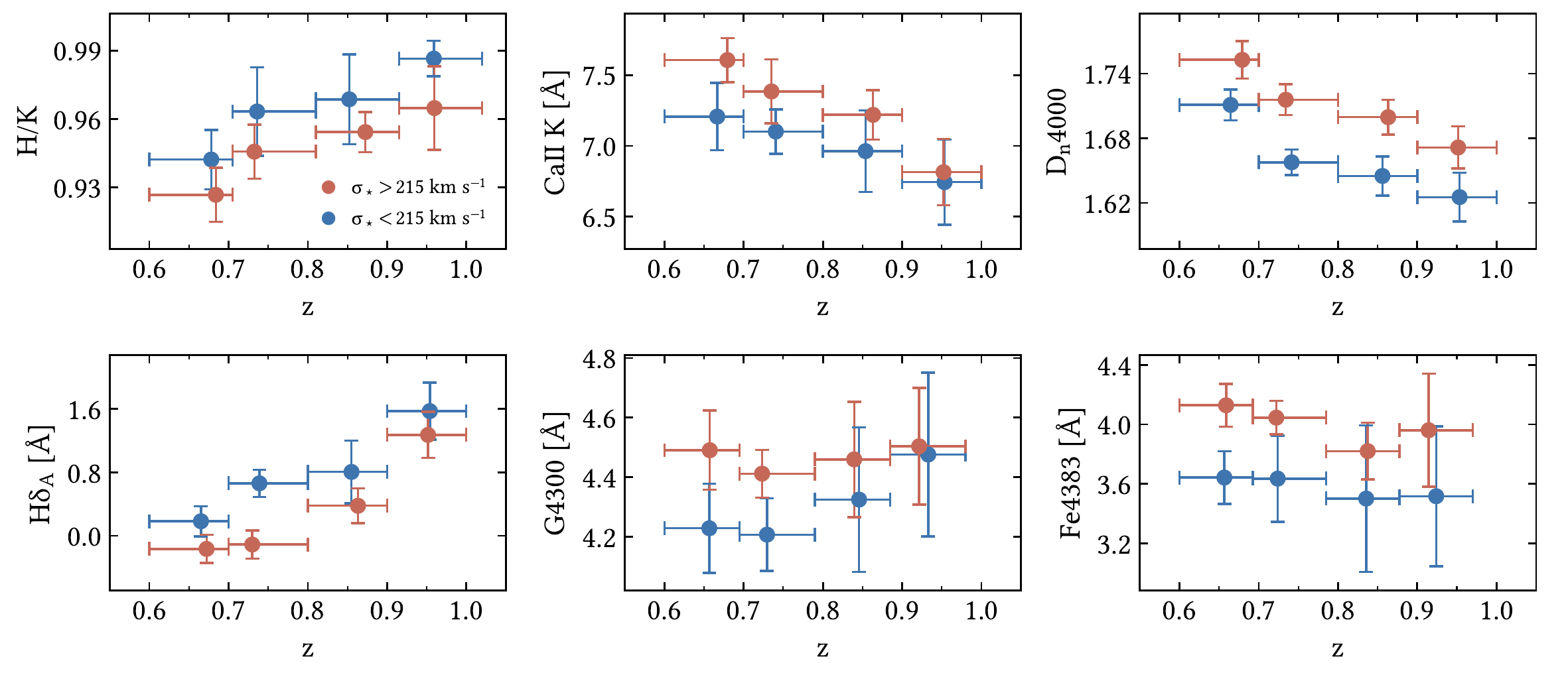}
      \caption{Binned relationships between the main spectral indices and redshift for the \textit{bona fide} passive sample divided into two velocity dispersion regimes ($\sigmastar=215$~\kms\ as threshold). Each bin contains on average $N\sim30$ galaxies. Error bars on the $x$-axis represent the bin width, while those on the $y$-axis are errors associated to the mean values.\label{fig:sample:ind_z_sigma}} 
    \end{figure*}

    We further explore the \textit{bona fide} passive sample by looking at index--$z$--\sigmastar\ trends. For this purpose, we firstly divide galaxies into two stellar velocity dispersion bins using $\sigmastar = 215$~\kms\ as the threshold value. Then, we divide each subsample into four or three redshift bins depending on the redshift coverage, considering intervals $\Delta z \sim 0.08-0.1$. All the bins have on average $N\sim 30$ objects and in all cases $N>10$. Finally, we compute the mean of the redshift and corresponding index in each bin, along with its associated errors. The resulting trends for the main spectral features are presented in Figure~\ref{fig:sample:ind_z_sigma}. We also quantify these trends within individual galaxies by using the nonparametric Spearman rank correlation test. 
    
    In general, we observe that Balmer indices correlate moderately with redshift with a Spearman $\rho\sim0.4$ ($p$-value $10^{-8}$). At fixed $z$, high-\sigmastar\ galaxies have weaker Balmer indices. Similar behavior is observed for the \ind{0} index and - with an opposite trend - for the H/K ratio. \ind{3} anticorrelates moderately with redshift with a Spearman $\rho\sim -0.3$ ($p$-value $10^{-5}$). At fixed $z$, there is a clear separation between the two \sigmastar\ regimes. Note that these relations could be (and have been) used in the cosmic chronometers framework to constrain the expansion history of the Universe, once the D4000--age relation is carefully calibrated \citep[][]{Moresco2011}. 
    Temporarily ignoring metallicity effects that will be discussed in the following, these observed trends are fully consistent with the mass-downsizing scenario, i.e. more massive galaxies formed earlier and in short timescales, and then experienced a passive evolution. Further support to this idea is given by iron indices, which are stronger for high-\sigmastar\ galaxies but do not show significant evolution over $\sim 2$ Gyr of cosmic time. This finding supports the scenario in which this population has already exhausted its gas reservoir, hence not being able to significantly evolve their metallic content, and is evolving passively as a function of cosmic time.
    
    The absence of a correlation in redshift for \ind{15} and \ind{9}, instead, shows that these indices cannot be used as age indicators throughout cosmic time. However, the segregation in \sigmastar\ is still consistent with the idea that they are sensitive to stellar population age in the very first Gyr after the formation. Another explanation is that the segregation is due to different stellar metallicities and abundances. Even in that case, they could be crucial to determine galaxy ages indirectly by breaking the age--metallicity degeneracy.
 
    It is also interesting to notice that \ind{12} shows a very flat behavior as a function of redshift, for both \sigmastar\ bins. This trend, confirmed also by the analysis of the stellar metallicity presented below, supports the scenario for which these systems formed the majority of their mass (and hence metallicity) at high redshifts and in very concentrated episodes of star formation, exhausting their gas reservoirs and therefore not being able to further change their metal content.
    
    Qualitatively similar trends are obtained also maximizing the number of galaxies per bin instead of using bins of constant $\Delta z$. Given the inhomogeneous redshift coverage of LEGA-C, this latter method helps to strengthen bin statistics but has the cost of losing leverage in redshift.

    \subsection{Morphology}
      We analyze the morphological content of the \textit{bona fide} passive sample using the Zurich Estimator of Structural Types (ZEST) classification \citep{Scarlata2007}, based on principal component analysis of the surface brightness profiles. We find that the great majority (71\%) of galaxies are classified as E/S0, 27\% as intermediate (where the contribution of the bulge is similar to that of the disk), 2\% (six galaxies) as irregular, and none show a late-type morphology. Among the six galaxies classified as irregular, only two appear in the final sample of constrained galaxies. Their morphology is not significantly disturbed and their spectra are typical of old stellar populations, therefore we still include them in the \textit{bona fide} passive sample and verify that their exclusion does not affect our results. Within the \textit{bona fide} passive sample, we observe no significant correlation between different morphological classes and \sigmastar. Interestingly, a similar percentage of E/S0 types (72\%) has also been found by \cite{Moresco2013} in their sample of $\sim 17000$ zCOSMOS galaxies. The two works share 127 galaxies (37\% of the \textit{bona fide} passive sample) and adopt the same morphological catalog. However, selection criteria are different because the authors make use of the best-fit SED matching a local E-S0 template and different color cuts, along with spectroscopic cuts based on both \OII\ and H$\alpha$ emission.

      The presence of passive systems with non-purely early-type morphologies has been already discussed in the literature \citep[e.g.,][]{Dressler1999,Pozzetti2010}. This result can be explained by considering the existence of a class of objects for which the morphological transformation lasts longer than changes in stellar population, i.e. galaxy colors redden before the galaxy reaches an early-type morphology. It is therefore improper to treat galaxies without an early-type morphology as contaminants.

      In conclusion, physical properties derived from SED fitting (Figure~\ref{fig:sample:selection_diagrams}) and observed spectral properties of the three different subsamples considered (Figures~\ref{fig:sample:composite_350_passive}, \ref{fig:sample:obs_prop_3x2}, and \ref{fig:sample:ind_z_sigma}), confirm the existence of a population of passive systems characterized by lower/absent star formation, older stellar populations, and higher metallicities with respect to the parent galaxy population, and consistent with the mass-downsizing scenario. These observational data also support that the \textit{bona fide} selection adopted is able to maximize the purity of the sample, providing a sample of massive and passive galaxies, with a negligible (if any) contamination by star-forming outliers. The analysis of each individual galaxy will add more granularity to this picture.

\section{Analysis}\label{sec:analysis}

With this work, we are exploring the capability of studying cosmic chronometers with Lick indices, which allow us to compare with several literature studies \citep[e.g.,][]{Thomas2010, Gallazzi2014, Onodera2015}. The stellar ages derived with this method are light weighted and can be biased toward lower values because young stars outshine the older ones \citep{Conroy2013a}, while metallicity and abundance are found to be less affected \citep{Serra2007}. When a young stellar population is present, mass weighted age - which can be derived with full spectral fitting codes - can provide a better estimate of the integrated galaxy SFH. From the analysis of the H/K ratio (Sect.~\ref{sec:CaHK}) we can exclude a contamination from young (200 Myr--1 Gyr) stars in our \textit{bona fide} passive galaxies, therefore also the difference between light- and mass-weighted is expected to be negligible.

To more quantitatively assess the robustness for the CC method, in Kang et al. (2022, \textit{in preparation}) we perform a full spectral fitting analysis using the publicly available BAGPIPES code \citep{Carnall2018} to determine the ages, metallicities, and SFHs of the \textit{bona fide} galaxies selected in this paper. While we refer to that paper for the detailed description of the methods adopted and results obtained, we anticipate here that the slope of the $\age-z$ relation is very robust and in perfect agreement between our two studies, with a percentage difference $\lesssim 5\%$ ($\lesssim0.2\sigma$).

  \subsection{Stellar population model}\label{sec:analysis:model}
    In this work, we adopt the \citet*{Thomas2011} models (hereafter \citetalias{Thomas2011}), which make predictions of the Lick indices by varying SSP-like stellar ages (\age), metallicities (\met), and $\alpha$-element enhancements (\afe) of the stellar population. These models have been widely used in the literature for similar Lick indices studies \citep[e.g.,][]{Johansson2012,Jorgensen2013,Onodera2015,Scott2017,Lonoce2020}. They are based on the evolutionary synthesis code of \cite{Maraston2005} and element response functions from \cite{Korn2005}, and are carefully calibrated with galactic globular cluster data. The main ingredients are single-burst SFH, \citet{Salpeter1955} initial mass function, MILES empirical stellar libraries \citep{SanchezBlazquez2006}, and \citet{Cassisi1997} stellar evolutionary tracks. In this work, we use the models provided at a MILES resolution of 2.5~\AA\ \citep{Beifiori2011}. 
        
    The original grid spans the following parameter space: $0.1<\age/\mathrm{Gyr}<15$, $-2.25<\met<0.67$, and $-0.3<\afe<0.5$ with a total of 480 grid points, each one corresponding to the prediction for a single SSP. We perform a three-dimensional linear interpolation of the grid to reach a resolution of $\Delta \age = 0.1$~Gyr, and $\Delta \met = \Delta \afe = 0.01$~dex. This process allows us to achieve higher precision in parameter estimation. We verified that our approach does not introduce any systematic bias, since by considering different grid choices the final results are always fully compatible with each other within $1\sigma$.

    These models assume an instantaneous burst of star formation. While this assumption might in principle lead to a significant underestimation of the global stellar age in mixed populations, we note here that our selection criteria were chosen to obtain a sample with a very concentrated SFH and to minimize the contamination from a significant residual star formation, as confirmed by the analysis of various indicators discussed in Section~\ref{sec:sample}. In particular, for this population we expect the SFHs to be extremely coeval, with very small durations ($\tau\lesssim 0.3$~Gyr, if modeled with a delayed exponential SFH), as confirmed from a parallel analysis \citep{Borghi2022a}.

  \subsection{MCMC Analysis}\label{sec:analysis:mcmc}
    To compare the measured absorption features to \citetalias{Thomas2011} models, we develop a fully Bayesian analysis pipeline. We assume that the uncertainties on indices are well-determined, Gaussianly distributed, and independent. A set of modeled indices, which are a function of parameters $\boldsymbol{\theta}=(\age, \met, \afe)$, can therefore be fitted to the observed ones using the log-likelihood function 
    \begin{equation}
      \ln \left(\mathcal{L}\right) = k - \frac{1}{2} \sum_i^{\rm N_{ind}} \left(\frac{I_i-I^{\mathrm{mod}}_i(\boldsymbol{\theta})}{\sigma_i}\right)^2,
    \end{equation}
    where $k$ is a constant, $I^\mathrm{mod}_i(\boldsymbol{\theta})$ the model prediction for the $i$th observed index $I_i$, and $\sigma_i$ its uncertainty. The posterior probability distributions of $\boldsymbol{\theta}$ are explored using the affine-invariant ensemble sampler \texttt{emcee} \citep{ForemanMackey2019}. Chains are initialized with 200 walkers randomly scattered around the center of the parameter space. Each walker performs at least 2000 model realizations. Parameters and uncertainties are defined as the median, and 16th and 84th percentiles of the marginalized posterior distributions. 
    
    We use flat priors that span the entire parameter space allowed from the models. An important point that we stress here is that we do not assume any cosmological prior for galaxy ages. This is a crucial point, and a difference with respect to other similar works, to avoid introducing cosmological biases in the age determination and keep the results cosmological-independent.

    After the analysis, we carefully assess the convergence of each chain. A detailed description is given in \ref{app:convergence}.
        
  \subsection{Index combination}\label{sec:analysis:combo}
    The choice of the index set to analyze must be carefully addressed because, given the number of the measured indices, there are more than one million possible combinations. The relative sensitivity of different indices to different stellar population parameters and abundances is not identical \citep{Tripicco1995,Korn2005,Lee2009}. Balmer and \ind{3} indices are better suited to constrain ages, Fe-dominated indices to measure the Fe abundance and total stellar metallicity, while Mg indices to estimate the $\alpha$ elements abundance. It is therefore intuitive that different index sets provide different constraints. 
    
    Because the ultimate goal of this work is to avoid any effect that could potentially introduce biases in the $\age-z$ relation, we choose to fit galaxies considering always the same set of indices. This means, on one hand, that using indices covering a wide redshift range will significantly reduce the number of galaxies analyzed; on the other hand, to maximize the number of analyzable galaxies the set of indices has to be chosen to span a small wavelength range (from 3600 to 4900~\AA; see the upper panel of Figure~\ref{fig:sample:composite_350_passive} and Table~\ref{tab:ind_prop}). The tradeoff between these two aspects combined has the cost of reducing by about one-third the number of galaxies measured but guarantees the crucial advantage of providing a homogeneous analysis. 
    
    \begin{deluxetable}{chhccc}
      \renewcommand{\arraystretch}{1.3}
      \tablenum{4}
      \tablecaption{Extract from the catalog of measured stellar ages, metallicities \met, and \afe and associated $16^\mathit{th}-84^{th}$ percentile uncertainties\label{tab:parameters_preview}}
      \tablewidth{0pt}
      \tablehead{\colhead{ID} & \nocolhead{RA} & \nocolhead{DEC} & \colhead{\age} & \colhead{\met} & \colhead{\afe} \\
        \colhead{[MMS2013]} & \nocolhead{} & \nocolhead{} & \colhead{Gyr} & \colhead{dex} & \colhead{dex} }
      \startdata
      133240 & 150 09 41.4 & 2 21 44.9 & $3.08_{-0.64}^{+0.64}$ & $+0.26_{-0.09}^{+0.12}$ & $+0.16_{-0.06}^{+0.06}$ \\
      133783 & 150 16 44.5 & 2 22 06.3 & $1.90_{-0.08}^{+0.07}$ & $+0.33_{-0.03}^{+0.05}$ & $+0.14_{-0.03}^{+0.03}$ \\
      134169 & 150 17 56.0 & 2 22 20.7 & $2.79_{-0.36}^{+0.36}$ & $+0.16_{-0.07}^{+0.08}$ & $+0.24_{-0.05}^{+0.05}$ \\
      139772 & 150 16 00.7 & 2 26 21.9 & $3.05_{-0.12}^{+0.14}$ & $+0.12_{-0.02}^{+0.02}$ & $+0.23_{-0.03}^{+0.03}$ \\
      205742 & 150 05 19.5 & 2 27 38.3 & $3.07_{-0.27}^{+0.39}$ & $-0.01_{-0.04}^{+0.06}$ & $+0.04_{-0.03}^{+0.03}$ \\
      206573 & 150 16 01.8 & 2 28 11.2 & $3.30_{-0.19}^{+0.29}$ & $-0.02_{-0.03}^{+0.02}$ & $+0.28_{-0.03}^{+0.03}$ \\
      207825 & 150 17 40.8 & 2 29 19.3 & $1.94_{-0.07}^{+0.12}$ & $+0.16_{-0.05}^{+0.01}$ & $+0.16_{-0.02}^{+0.02}$ \\
      \multicolumn{6}{c}{\dots} \\
      \enddata
      \tablecomments{Full table is available online.}
    \end{deluxetable}

    For these reasons, we decide to use the following set of Lick indices: \ind{4}, \ind{6}, \ind{7}, \ind{8}, \ind{9}, \ind{10}, \ind{11}, \ind{12}, \ind{14}, and \ind{15}. They are chosen among those that are calibrated against globular cluster data in \citetalias{Thomas2011}, but excluding those redder than \ind{16}, because, as mentioned above, they would not allow us to obtain a statistically meaningful sample. We also exclude \ind{16} because it can be biased by the presence of a residual emission component \citep[e.g.,][]{Concas2017}. The final set spans a narrow wavelength range in the optical regime, from 4000 to 4800~\AA. We verified on a few available galaxies that the inclusion of \ind{20} does not significantly change the results. We also performed extensive tests with very different sets of indices. A detailed discussion is presented in \ref{app:combos}. Given the available data, we find that this is the optimal set, because it maximizes both the spectral coverage and the number of galaxies for which we obtain constraints, and it also provides a good balance between age-, metallicity-, and $\alpha$-sensitive indices.

    In the end, we obtain robust constraints for 140 galaxies over 199 that have been analyzed (Table~\ref{tab:parameters_preview}), after removing the galaxies with nonconverging fits (see \ref{app:convergence}). These galaxies are at $\med{z}=0.70$, and have spectral $\med{\rm S/N}=26.4$ per pixel. Typical uncertainties are $\pm 0.33$ Gyr in \age and $\pm 0.05$ dex in \met and \afe.

\section{Results and discussion}\label{sec:res}
  In this section, we present the results obtained from the analysis of the previously defined H/K ratio and for the physical properties of selected passive galaxies in the LEGA-C DR2 survey. We start by discussing the H/K ratio and the correlation with commonly used diagnostics in Section~\ref{sec:res:HK}. In Section~\ref{sec:res:parsigma} we explore scaling relations of stellar population parameters versus \sigmastar\ and \mstar. In Section~\ref{sec:res:parz} we discuss trends with redshift, with a focus on the age--redshift relation. Finally, in Section~\ref{sec:res:binned} we present the median binned parameters--redshift relations.  
  
  We underline here below the main points of our analysis to be kept in mind when comparing our results with those of previous analyses in the same field.
  
  \begin{enumerate}[(i)]
    \item Our galaxies are passive. Similar studies are mostly targeting morphological early-type galaxies. These samples may contain galaxies with a nonnegligible level of star-formation ($\sim 20\%$ at the current stellar masses; \citealt{Moresco2013});
    \item We do not use cosmological priors. We also verified that, especially for low-S/N galaxies, the use of a cosmological upper limit for the galaxy ages may produce an apparent convergence of Markov Chain Monte Carlo (MCMC) chains at higher ages toward the prior, introducing a bias in the sample (\ref{app:convergence});
    \item We adopt single-burst SFHs. In this case, the time of the onset of star-formation $t_\mathrm{form}$ coincides with the time of the main (and only) star-formation event $t_{\rm peak}$, and with the time of quenching $t_{\rm quench}$. A proper comparison with other data sets where an extended SFH is assumed should carefully take into account how ages are defined and intrinsic degeneracies between SFH parameters. For instance, by assuming an $\mathrm{SFR}(t)\propto \exp( (t-\age)/\tau)$ (tau model), a positive correlation between the timescale $\tau$ and the galaxy age (time since $t_\mathrm{form}$) can be present, with values of $\Delta \tau \sim 0.3~\Delta \age$ \citep{Borghi2022a}.
  \end{enumerate}

  Before applying the cosmic chronometer method, points (i) and (ii) should be carefully considered, but it is important to stress that this method relies on differential - not absolute - galaxy ages. Therefore, by assuming a more extended SFH for the entire population of these galaxies (i.e. a vertical offset in the $\age-z$ plane), the final $H(z)$ measurement is not affected (see Section~\ref{sec:analysis:model}).

  \begin{figure*}
    \centering
      \includegraphics[width=0.95\hsize]{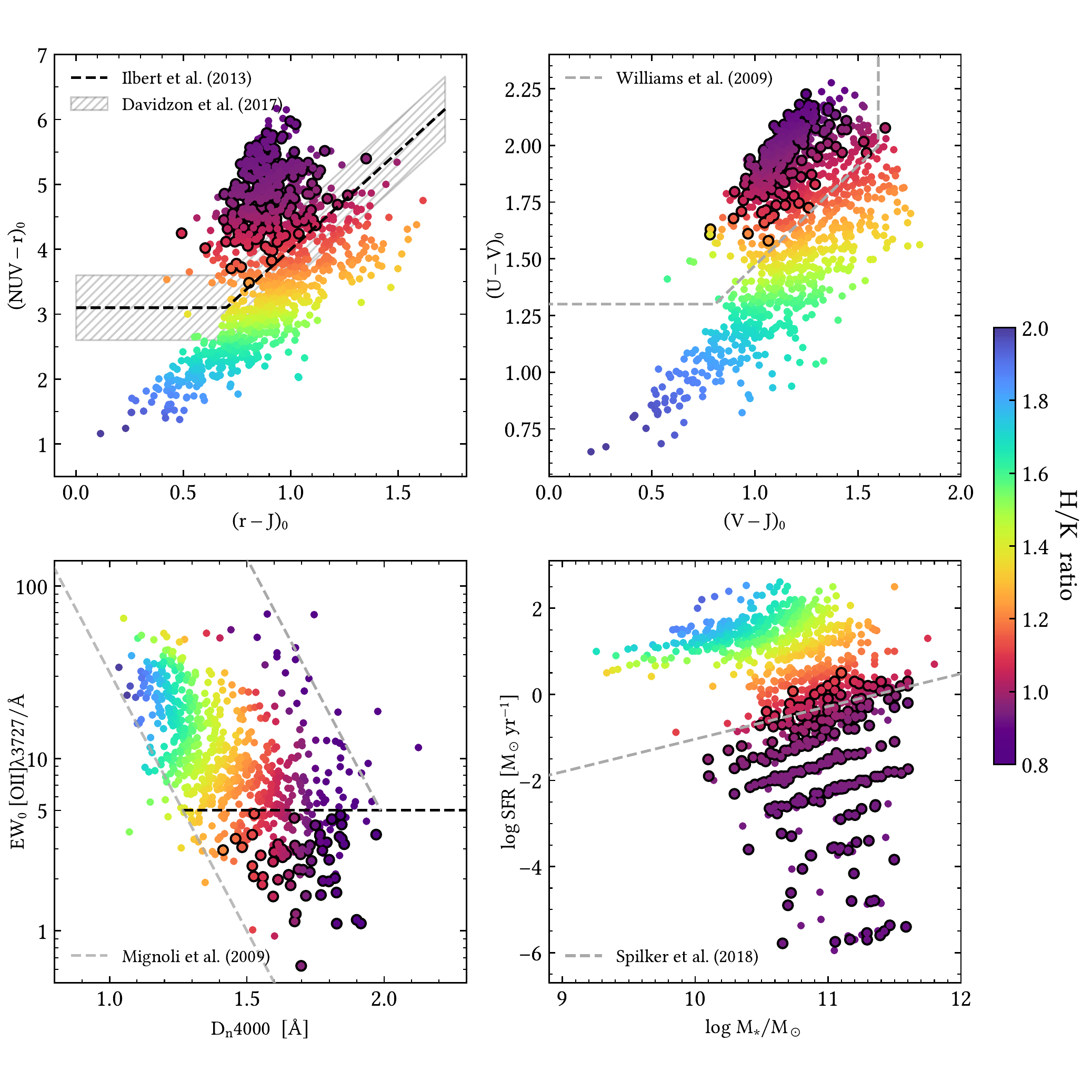}
      \vspace{-1em}
      \caption{Four well-known diagnostic diagrams (NUVrJ, UVJ, \EWOII--\ind{3}, SFR--\mstar) color-coded by \ind{24} using the LOESS method (see text). Each panel shows the \textit{parent sample} galaxies extracted from LEGA-C DR2 with a S/N of the \ind{24} higher than 3. Black borders identify \textit{bona fide} passive galaxies. Dashed lines are the criteria to separate star-forming from passive galaxies taken from the literature \citep[][respectively]{Ilbert2013,Williams2009,Mignoli2009,Spilker2018}. We use black colors for the criteria also adopted in this work, while gray lines are for illustrative purposes only. \label{fig:res:dignostic_HK_4x}}
  \end{figure*}

  \subsection{The H/K ratio as a stellar population diagnostic}\label{sec:res:HK}
    In Figure \ref{fig:res:dignostic_HK_4x} we show the distribution of the entire parent sample in four widely used diagnostic diagrams, namely NUVrJ, UVJ, \EWOII--\ind{3}, and SFR--\mstar, color-coded by the H/K ratio. To better capture the mean trends of H/K, we perform locally weighted regression (LOESS). We use the \texttt{LOESS} package of \cite{Cappellari2013b} based on the two-dimensional algorithm of \cite{Cleveland1988}, with a linear local approximation and a regularization factor $f=0.5$. Selected \textit{bona fide} passive galaxies are highlighted with black borders. To quantify these trends, we use Spearman correlation coefficients.

    The diagrams show a clear and significant correlation between the H/K ratio and other diagnostic tools despite its small dynamic range ($0.8\lesssim {\rm H/K} \lesssim 2$, and a median error of $\sim 0.14$). In more detail, we observe a strong correlation with (NUV--$r$) ($\rho = -0.72$; $\text{$p$-value}=10^{-164}$) and with ($U-V$) ($\rho = -0.72$; $\text{$p$-value}=10^{-178}$). The correlation with ($r-J$) and ($V-J$) colors is weaker but still significant ($\rho = -0.25$; $\text{$p$-value}=10^{-19}$ and $10^{-15}$, respectively). Interestingly, we find that a selection based on the threshold value $\rm{H/K}<1.1$ can reproduce the NUVrJ selection with $17\%$ incompleteness and $19\%$ contamination. 
    
    As shown in the third panel, an H/K cut does not exclude galaxies with a significant \OII\ emission. For instance, $19\%$ of the plotted galaxies have ${\rm H/K}<1$ and $\EWOII>5$~\AA. Unfortunately, the lack of other spectral features as H$\alpha$, [N \textsc{ii}], and [S \textsc{ii}], due to the limited wavelength coverage of the current data set, does not allow us to investigate the nature of these sources. For these systems, the combination of multiple indicators is still needed to obtain a pure sample of passive galaxies. 

    A strong correlation is also observed for the sSFR ($\rho=0.70$; $\text{$p$-value}=10^{-161}$). We find that a threshold on H/K values of ${\rm H/K}<1.1$ can reproduce a ${\rm \log sSFR/yr}<-11$ cut with $15\%$ incompleteness and $16\%$ contamination. 
    
    These are remarkable results if we compare the data needed for the two different selections, and the range of wavelengths spanned. On one hand, a wide photometric coverage is needed for a reliable estimate of a NUVrJ diagram, SFR, or \mstar\ (typically from the UV to the near-IR), with an accuracy increasing with the number of available photometric points; it is therefore not always available in many surveys. On the other hand, we have a feature defined over a window of only about $150$~\AA\ for which deep rest-frame optical spectroscopy is needed. The H/K can therefore play a key role in the selection of pure samples of passive galaxies in future wide-field grism surveys such as Euclid \citep{Laureijs2011} and the Roman Space Telescope \citep{Spergel2015}. Another advantage of this diagnostic is the mild dependence on spectral resolution. Differences between H/K values obtained on 8 and $2.5$~\AA\ FWHM spectra are $\lesssim 4\%$. Performing the same analysis on individual \ind{0} and \ind{1} indices yields $\sim 10\%$ systematic differences.

    \begin{figure}[t]
      \includegraphics[width=\hsize]{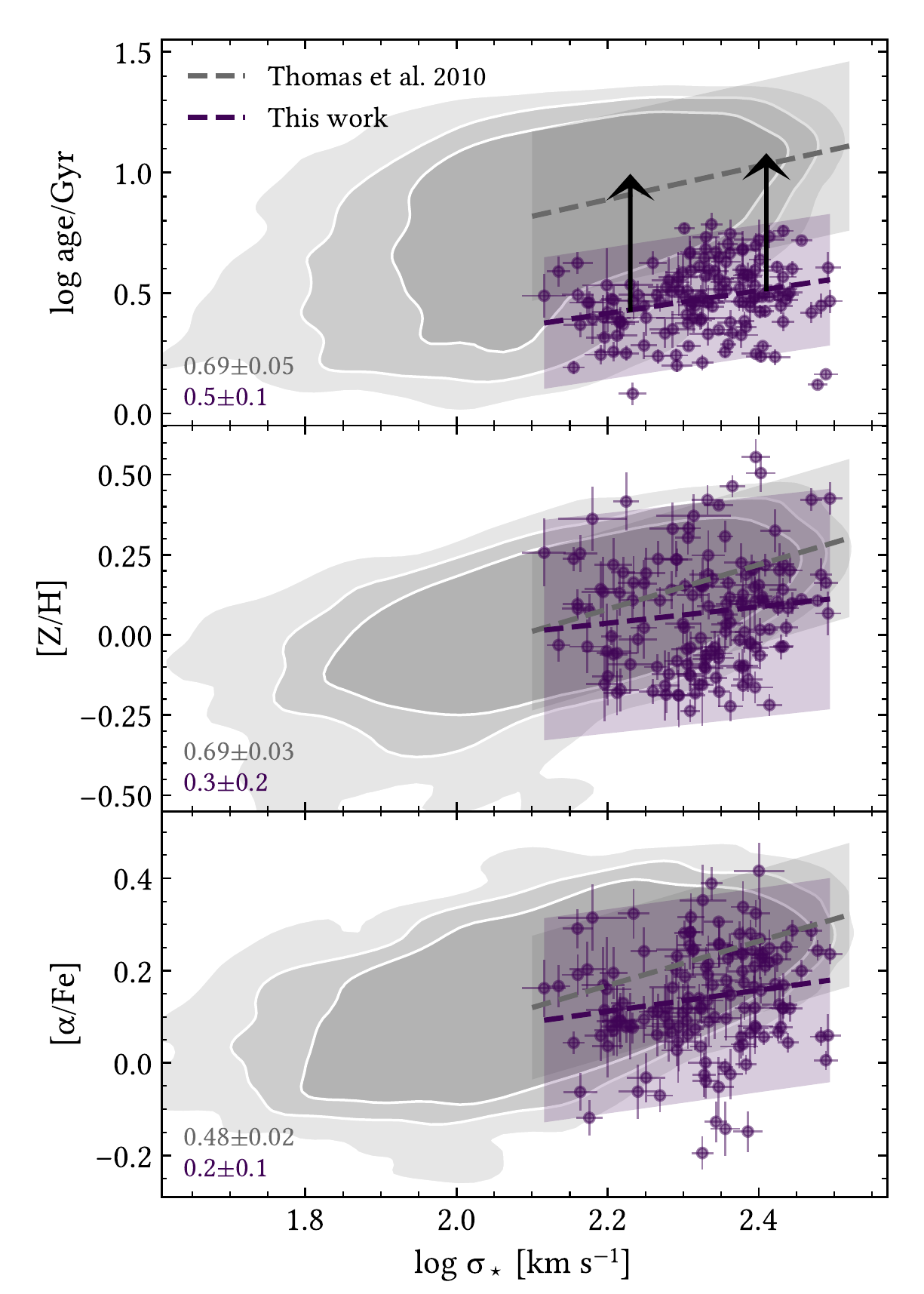}
      \caption{Distribution of mean stellar ages, metallicities, and $\alpha$ enhancements as a function of stellar velocity dispersion for individual massive and passive galaxies selected in LEGA-C DR2 (violet points). Gray contours represent early-type galaxies in the local Universe from SDSS/MOSES \citep{Thomas2010}, and enclose $1\sigma$, $2\sigma$, and $5\sigma$ regions. Dashed lines and shaded regions are robust linear fits and associated $2\sigma$ scatter regions, for $2.1<\log\sigmastar<2.5$. The resulting slopes and their uncertainties are annotated in the bottom left. Black arrows represent the age evolution expected for a passive SSP.\label{fig:res:parsigma}}
    \end{figure}

    Concerning the 140 \textit{bona fide} passive galaxies, we note here that no correlation between stellar population parameters (especially age) and H/K is present. This result is expected given the fact that no significant contribution from a young stellar component is present, i.e. the H$\epsilon$ line deepening effect becomes negligible in the passive regime.
    
    In summary, young ($\sim 200$ Myr) stellar populations whose light is predominantly due to A- and B-type stars (${\rm H/K} > 1.1$) are characterized by higher UV fluxes, lower D4000, a higher \EWOII, and dominate the star formation main sequence of LEGA-C galaxies. The H/K lowers when approaching the quiescence criteria, but the study of timescales and the interplay with stellar population parameters for the whole population of galaxies will require a further assessment.

    \subsection{Physical parameters versus $\sigmastar$ and $\mstar$}\label{sec:res:parsigma}
      Figure~\ref{fig:res:parsigma} shows stellar population parameters as a function of the observed stellar velocity dispersion for passive galaxies at $z\sim 0.7$. As a local control sample, we consider the SDSS/MOSES data set of \cite{Thomas2010} of morphologically selected early-type galaxies (ETG) at $z\sim 0.05$, since similar models and analysis techniques are adopted. Their sample is $\sim 20$ times larger than ours and spans a wider range in $\log \sigmastar$ (1.7--2.5). For a fair comparison, the local sample is limited to comparable stellar velocity dispersion values, i.e. $2.1<\log\sigmastar~[\kms]<2.5$. To this catalog we add galaxy stellar-mass estimates from MPA-JHU DR 8 \citep{Kauffmann2003}. To study $\log\age$, \met, and \afe versus $\log \sigmastar$ and $\log M_\ast$ we perform robust linear regression with the least trimmed squares (LTS) algorithm \citep{Rousseeuw1984} and measure their Spearman coefficients. Results are quoted in Table~\ref{tab:scaling_rel}.

      \begin{deluxetable}{ccccc}
        \renewcommand{\arraystretch}{1.3}
        \tablenum{5}
        \tablecaption{Coefficients for the Scaling Relations in the \textit{bona fide} Passive Sample and Associated Spearman Coefficients and $p$-values\label{tab:scaling_rel}}
        \tablewidth{0pt}
        \tablehead{\colhead{y} & \colhead{$a \pm {\rm err}(a)$} & \colhead{$b \pm {\rm err}(b)$} & \colhead{rms} & \colhead{$\rho$ ($p$-value)}
          }
        \startdata
        \multicolumn{5}{c}{$y = a\;\log\sigmastar + b$} \\
        $\log\age/\mathrm{Gyr}$ & $0.5 \pm 0.1$ & $-0.6 \pm 0.3$ &  $0.1$ & $0.2$ $(0.02)$ \\
        \met & $0.3 \pm 0.2$ & $-0.5 \pm 0.4$ &  $0.2$ & $0.2$ $(0.15)$ \\
        \afe & $0.2 \pm 0.1$ & $-0.4 \pm 0.3$ &  $0.1$ & $0.2$ $(0.03)$  \\
        \hline
        \multicolumn{5}{c}{$y = a\;\log(M_\star/10^{11} M_\odot) + b$} \\
        $\log\age/\mathrm{Gyr}$ & $0.19 \pm 0.04$ & $0.47 \pm 0.01$ &  $0.1$ & $0.3$ $(<0.01)$ \\
        \met & $0.03 \pm 0.05$ & $0.07 \pm 0.02$ & $0.2$ & $0.0$ $(0.63)$ \\
        \afe & $0.02 \pm 0.03$ & $0.14 \pm 0.01$ & $0.1$ & $0.1$ $(0.50)$ \\
        \enddata
        \tablecomments{Linear fits are obtained with the \texttt{LtsFit} routine \citep{Cappellari2013a}.}
      \end{deluxetable}
  
      \begin{figure}[t]
        \centering
        \includegraphics[width=\hsize]{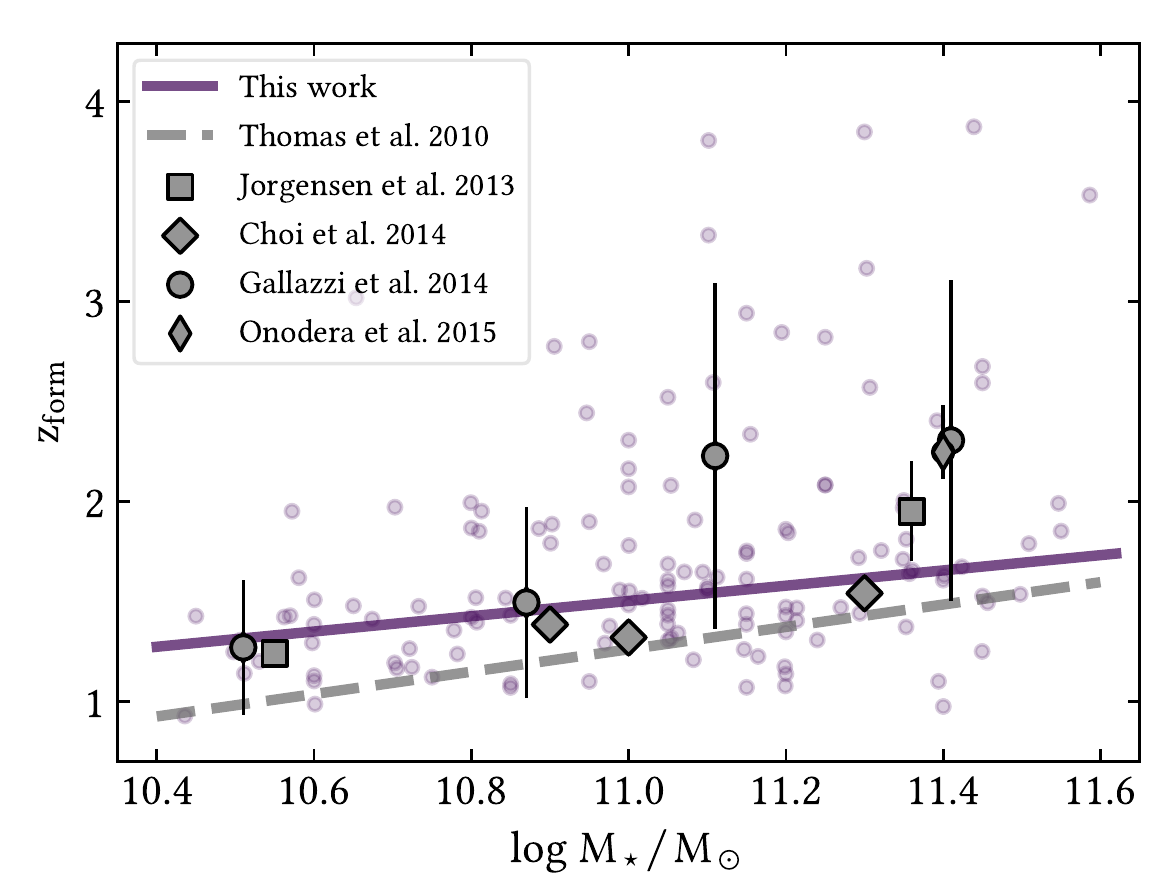}
            \caption{Formation redshifts as a function of galaxy stellar mass for 140 \textit{bona fide} passive galaxies selected in LEGA-C DR2 (violet points). We compare our measurements with literature data of massive quiescent galaxies in the local \citep{Thomas2010} and intermediate-redshift \citep{Jorgensen2013,Choi2014,Gallazzi2014,Onodera2015} Universe. Lines are obtained by performing robust linear fits (see text).\label{fig:res:zf_mass}}
      \end{figure}

      The general trends in \sigmastar\ are consistent with a stellar population that experienced a passive evolution from $z\sim 0.7$ to $z\sim 0.05$. In our $z\sim0.7$ sample we find shallower relations between the stellar population parameters and \sigmastar. This is expected from the lower statistics and the different selection criteria adopted. Moreover, the dynamic range available in \age\ decreases with increasing redshift because the Universe gets younger. 
      
      We first consider SSP-equivalent stellar ages. Remarkably, the difference of 5.5~Gyr between the two samples is perfectly in agreement with the age evolution of the Universe, confirming that the stellar populations of this population of galaxies experienced a pure passive evolution within the assumed reference cosmology. Contamination from young stellar populations in the high-$z$ (low-$z$) sample would produce a larger (smaller) offset. We find mild correlations between $\log\age$ vs. $\log\sigmastar$ (with a slope of $0.5 \pm 0.1$) and $\log\age$ vs. $\log\mstar$ (with a slope of $0.19 \pm 0.04$). To facilitate the comparison between different works, we convert galaxy ages to formation redshifts $z_\mathrm{form}$, and study the relation with \mstar, obtaining
      \begin{equation}\label{eq:zf}
        z_\mathrm{form} = \left(0.40 \pm 0.05\right) \: \log_{10} \left(\frac{\mstar}{10^{11}\, M_\odot}\right) + \left(1.46 \pm 0.02\right),
      \end{equation}
      with an intrinsic scatter of $\sim 0.24$. This means that galaxies with higher stellar mass ($\log\mstar/M_\odot=11.3$) formed their stars at $ z_\mathrm{form}\sim 1.6$, while less massive ones ($\log\mstar/M_\odot=10.7$) formed their stars at $z_\mathrm{form}\sim 1.3$. Interestingly, the formation epoch of this population of passive galaxies is found to take place right after the peak of the cosmic star formation rate density \citep[$z\sim 2$;][]{Madau2014}. We also find few (22, $\sim 16\%$ of the passive sample) very massive ($\log\mstar/M_\odot>11$) galaxies with a $z_\mathrm{form}> 2.5$ up to 5.
    
      In Figure~\ref{fig:res:zf_mass} our results are compared with the existing literature where similar analysis methods are used. We find very good agreement with the work of \cite{Jorgensen2013}, who studied spectra of $\sim 80$ cluster galaxies at $z=0.5-0.9$ comparing observed Lick indices with \citetalias{Thomas2011} models. They found formation redshifts of $z_\mathrm{form}\approx 1.24$ and $1.95$ for stellar masses of $\log\mstar/M_\odot\approx 10.6$ and $11.4$, respectively. We also find an excellent agreement with the work of \cite{Choi2014}, who analyzed stacked spectra of sSFR-selected passive galaxies at comparable redshifts and masses with the full spectral fitting method. By assuming single-burst SFHs, the authors found typical formation epochs of $z_\mathrm{form}\sim 1.5$. Finally, \cite{Gallazzi2014} analyzed ages and stellar metallicities for $\sim 70$ between star-forming and quiescent galaxies at $z\sim 0.7$ using age-sensitive \ind{3}, \ind{16}, and \ind{4}+\ind{10}, and metal-sensitive ${[\rm Mg_2Fe]}$ and ${\rm [MgFe]^\prime}$ indices. Rewriting Equation~\ref{eq:zf} in terms of the formation time $t_\mathrm{form}$ versus $\log\mstar$, we obtain a trend of $-1.26\pm0.27$~Gyr per decade in mass. Remarkably, this is in excellent agreement with the trend observed by \cite{Carnall2019} of $-1.48\pm0.37$~Gyr per decade in mass, using a very different approach (full spectral fitting) and assuming more extended (double power-law) SFHs.

    We now move to the analysis of mean stellar metallicities. Our results show no significant evolution in \met for passive galaxies since $z\sim 0.7$, with a median offset of 0.05~dex comparable to the median uncertainty. This confirms and statistically strengthens earlier results from \citet{Gallazzi2014} at similar redshifts, and it confirms several works that found similar metallicities up to $z\approx 2$ \citep{Onodera2012,Onodera2015,Citro2016,EstradaCarpenter2019}. On the other hand, recent studies report a significant evolution of $\Delta \rm{[Z/H]}>0.1$~dex. An example is the work by \citet{Beverage2021} based on a sample of 65 LEGA-C quiescent galaxies analyzed with a full spectral fitting code. The authors found no evolution in $\rm{[Mg/Fe]}$ values, but a $\rm{\Delta[Fe/H]}$ (hence $\rm{\Delta[Z/H]}$) of about 0.2~dex with respect to local $\log(\mstar/M_\odot)=11$ quiescent stacks. We also study \met versus $\log\sigmastar$ and $\log\mstar$, finding no significant correlations. Again, this can be attributed to our stricter selection criteria of the most passive systems, which are known to have shallower metallicity--mass relations \citep[see][]{Peng2015,Gallazzi2014,Gallazzi2021}. Indeed, when dividing them into two \sigmastar\ bins, we find that the typical \met\ of $\sigmastar>215~\kms$ systems is 0.1~dex higher than low-$\sigma$ ones, consistently with the downsizing scenario. 

    \begin{figure*}[t]
      \centering
      \includegraphics[width=0.49\hsize]{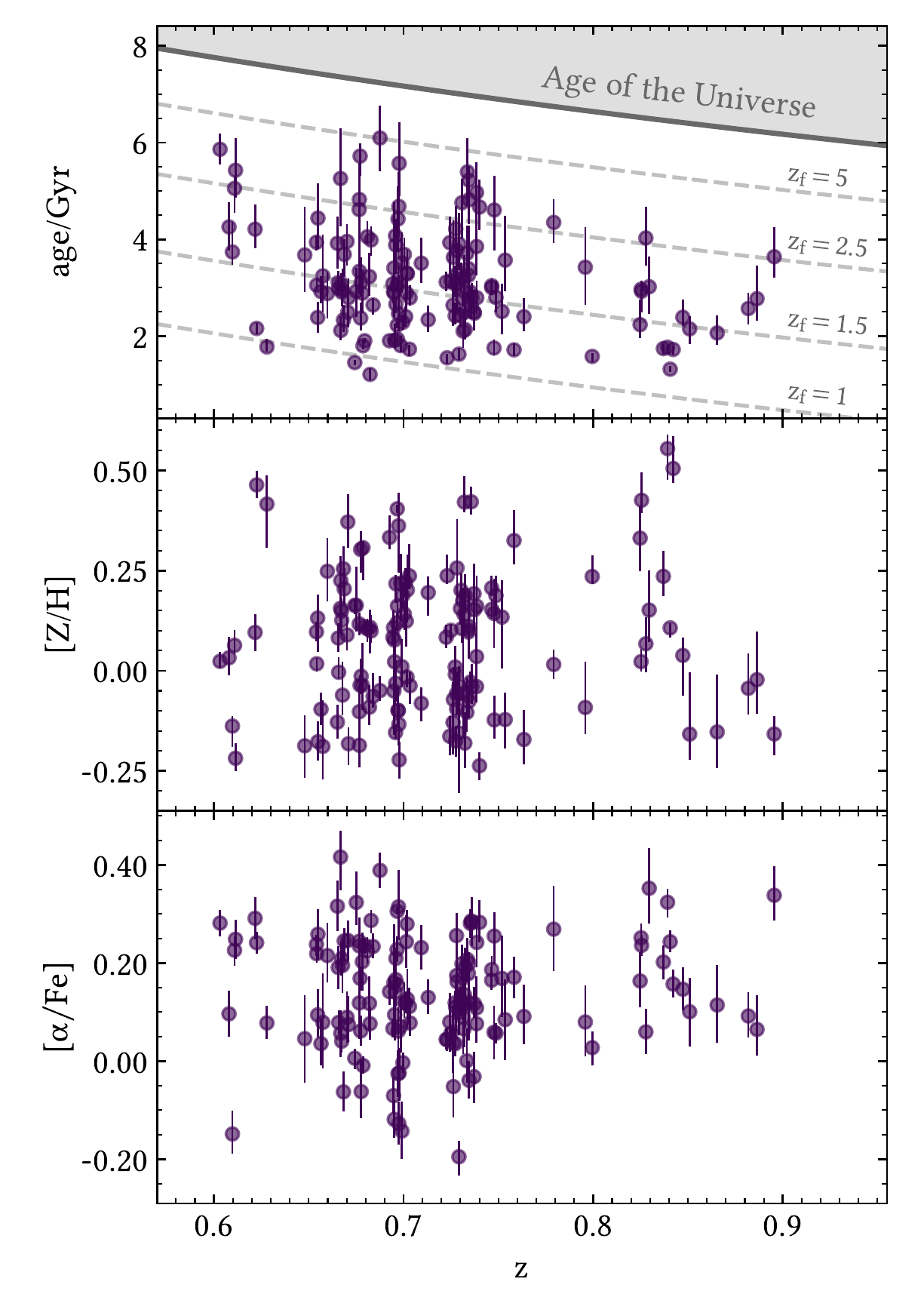}
      \includegraphics[width=0.49\hsize]{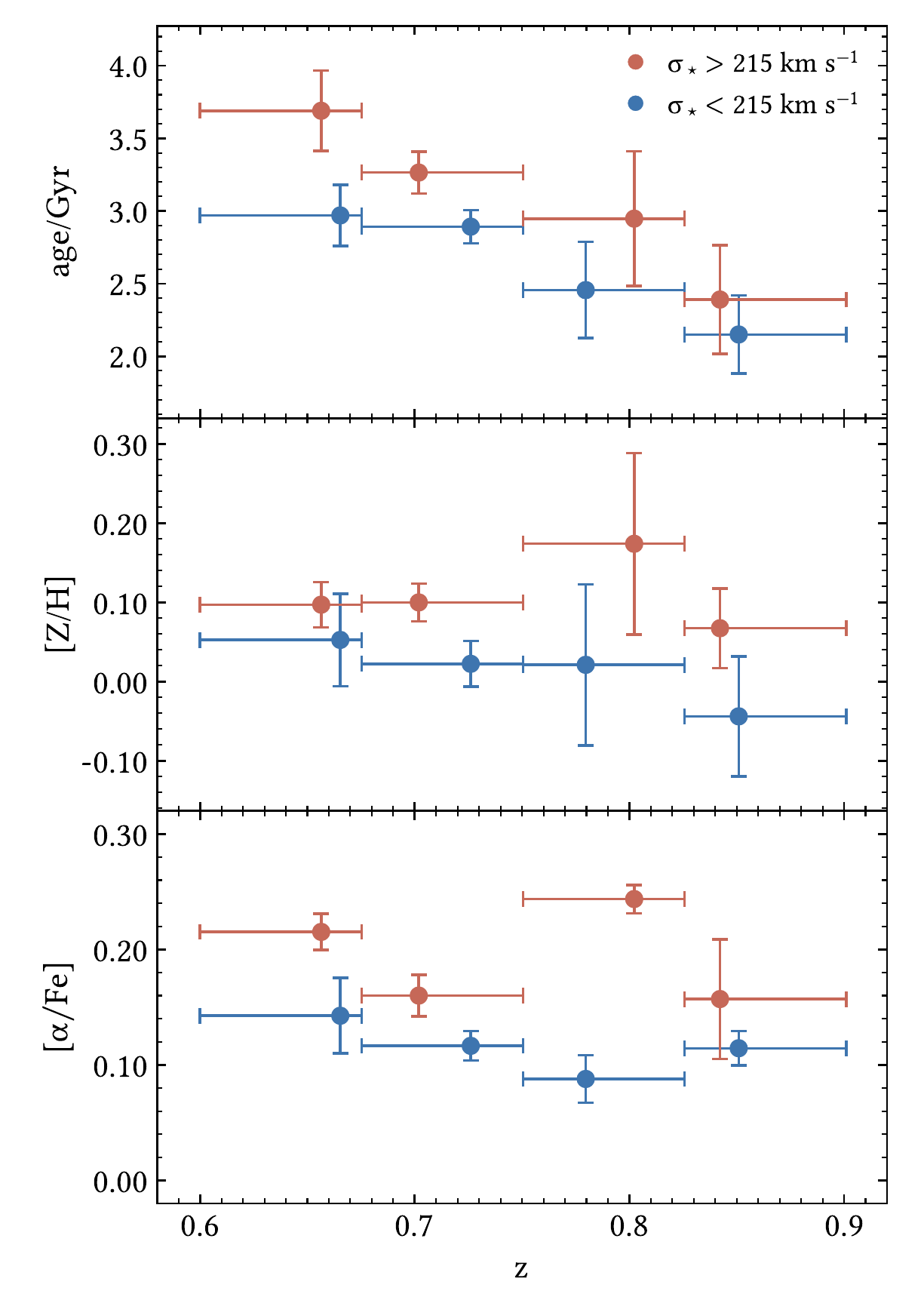}
          \caption{Left panels: distribution of single-burst stellar ages, metallicities, and $\alpha$-enhancements as a function of redshift for 140 \textit{bona fide} passive galaxies selected in LEGA-C DR2 (violet points). The error bars are obtained as the 16$^{th}$ and 84$^{th}$ percentiles of the marginalized posterior distributions. In the age--redshift panel, we shade the parameter space not allowed for a general cosmology (gray solid), as well as the formation redshift assuming a pure passive evolution (dotted lines). Right panels: median binned relations obtained by dividing the sample into two \sigmastar\ regimes, with $\sigmastar=215$~\kms\ as a threshold. Each bin contains 5--40 objects. Error bars on the $y$-axis are errors associated to the median values, while those on the $x$-axis indicate the bin widths.\label{fig:res:param_redshift}}
    \end{figure*}

    Finally, we analyze the mean stellar $\alpha$ enhancements. The \afe\ values of our galaxies are $\sim0.1$~dex lower than the local sample. However, as discussed in \ref{app:combos}, this could be completely explained by the fact that we could not use Mg indices in our baseline analysis. For the first time, we identify a positive trend between \afe\ and $\log\sigmastar$ with a slope of ($0.2\pm0.1$) for a large population of individual passive galaxies at $z\sim0.7$. This confirms the trends observed with stacks of quiescent galaxies at $0.1<z<0.7$ by \cite{Choi2014}. We do not find a significant trend in $\log\mstar$. On one hand, local early-type galaxies show stronger correlations with the gravitational potential well (of which \sigmastar\ is a tracer) than with \mstar\ \citep[e.g.,][]{Thomas2005,Barone2018}. On the other hand, uncertainties in the \mstar\ derivation may play a role in washing out these relations.

    In conclusion, the sample of selected passive galaxies at $z\sim 0.7$ shows trends in \age, \met, and \afe versus \sigmastar\ in agreement with those expected from a passively evolving population. The age offset, as well as the lack of a significant offset between the typical values of \met and \afe is evidence that these systems should have formed their stars on short timescales, depleting the great majority of their gas reservoirs, and experienced passive evolution since then. 
  
  \subsection{Physical parameters versus $z$}\label{sec:res:parz}
    Figure~\ref{fig:res:param_redshift} shows the derived stellar population parameters as a function of redshift. Interestingly, despite not having imposed any cosmological prior on the age of the galaxies, our derived ages are in all cases in agreement with a generic cosmological model, never exceeding the age of the Universe at any redshift. Moreover, the upper envelope of the distribution follows the decrease expected from the aging of the Universe. At all redshifts, only a few passive galaxies have ages $\lesssim 2$~Gyr owing to the stringent selection criteria. The median value is $\med{\age}=3.01$~Gyr with a $\pm 0.97$~Gyr  $1\sigma$ scatter. This implies a formation time of $t_\mathrm{form}\sim 4$~Gyr after the Big Bang, corresponding to a formation redshift of $z_\mathrm{form}\sim 1.5$ as previously discussed (Section~\ref{sec:res:parsigma}).

    Stellar metallicities have solar or slightly supersolar values, $\med{\met}=0.08$ dex with a $\pm 0.18$  $1\sigma$ scatter, spanning a very narrow range if compared to the initial parameter space ($-2.25<\met<0.67$). Differently from stellar ages, they show no sign of evolution even within the redshift range explored in this work.

    We find slightly supersolar \afe values, $\med{\afe}=0.13$ dex with a $\pm 0.11$  $1\sigma$ scatter. In particular, 124 (89\%) galaxies have $\afe > 0$, pointing to very short formation timescales, i.e. before Type Ia supernova explosions can pollute the interstellar medium with a relatively high amount of iron-peak elements. As for the metallicities, these star formation timescales indicators do not significantly evolve over the redshift probed.

    Overall, these results confirm the passive evolution within the $\sim1.6$~Gyr interval of cosmic time explored in this work shedding light on the granularity of the physical properties and SFHs. The uniformly small scatter in \met and \afe, $\lesssim 0.20$~dex at fixed \sigmastar\ or $z$, confirms the large homogeneity of the sample and puts strong constraints on the duration of the chemical assembly of these systems.

  \subsection{Median binned relations}\label{sec:res:binned}
    To provide a comprehensive picture with the results discussed in Section~\ref{sec:res:parsigma} and \ref{sec:res:parz}, we bin \age, \met, and \afe in $\sigmastar$ and in $z$. Binning in \sigmastar\ instead of \mstar\ has the benefit of avoiding model-dependent effects introduced by SED-fit modeling, other than using an observational quantity. The galaxies were firstly binned into two \sigmastar\ using $\med{\sigmastar}=215$~\kms\ (approximately equivalent to $\med{\log\mstar/M_\odot}\approx 11$) as a threshold, then into four equally spaced redshift bins, with $\Delta z = 0.075$ from $z=0.6$ to $z=0.9$. Bins have from $\sim 5$, at higher $z$, to $\sim 40$, at lower $z$, objects. To each bin, we assign a mean $z$ value and a median (\age, \met, \afe) with associated uncertainty. Results are shown in the right panels of Figure~\ref{fig:res:param_redshift}.

    Clearly, the median properties of the analyzed passive galaxies follow a downsizing pattern. At each cosmic epoch, stellar populations hosted in galaxies with higher mass are older, more metal rich and $alpha$ enhanced. This suggests that their formation occurred at earlier times, with a difference of $\Delta\age\approx 0.5$~Gyr, and on shorter time scales with respect to less massive ones. We note that these trends were already qualitatively confirmed from the analysis of the main absorption indices (Section~\ref{sec:sample:obs_prop}).
    
    Last but not least, it is remarkable that for each \sigmastar\ regime we find a clear, almost parallel, age--redshift relation. The study of their differential evolution will allow us to perform cosmological studies using the cosmic chronometer approach \citep{Borghi2022a}. 

\section{Conclusions}\label{sec:conclusions}
  In this work, we take advantage of the public Data Release 2 of the LEGA-C spectroscopic survey to place constraints on the stellar population properties of individual massive and passive galaxies at $0.6<z<0.9$. Based on a robust spectral analysis of Lick indices, our aim is to characterize this population and to explore the reliability of using these galaxies as cosmic chronometers. Our main results are summarized below.

  \begin{enumerate}\itemsep3pt
    \item We select a pure sample of 350 passive galaxies at $z\sim 0.7$ combining a photometric NUVrJ selection, a spectroscopic \EWOII\ cut, and a careful visual inspection of individual spectra to further remove galaxies with significant emission-lines (Figure~\ref{fig:sample:selection_diagrams}). As confirmed by the stacked spectrum (Figure~\ref{fig:sample:composite_350_passive}), no underlying emission line components are present in the sample, confirming its high purity. Selected passive galaxies have a median observed velocity dispersion of $\med{\sigmastar}=206$~\kms, stellar mass of $\med{\log\mstar/M_\odot}=10.95$, and very low specific star formation rate $\med{\rm \log sSFR/yr}=-12.1$. Most of them have an early-type morphology, but there is also a nonnegligible percentage of systems (about one-third) with an intermediate morphology.
  
    \item We develop, validate, and publicly release \texttt{PyLick}, a flexible Python tool to measure absorption features, implementing several different index definitions (see \ref{app:pylick}). This allows us to measure spectral indices over a wide wavelength range in LEGA-C data, extending the current public catalog of Lick indices by \cite{Straatman2018} and enabling a more detailed exploration of the dependence of our results on different index combinations.
    
    \item We introduce the H/K ratio, a new diagnostic feature defined as the ratio of pseudo-Lick indices \ind{0} and \ind{1} (Figure~\ref{fig:sample:HK_definition}). We verify that it is an excellent tracer of potential contamination of the sample due to star-forming or young populations, confirming that our sample is compatible with no or negligible contamination, with $\med{\rm H/K}=0.96\pm 0.08 ~(1\sigma)$. Moreover, a selection based on ${\rm H/K}<1.1$ is found to reproduce a NUVrJ selection \citep{Ilbert2013} or a ${\rm \log sSFR/yr}<-11$ cut (Figure~\ref{fig:res:dignostic_HK_4x}) with a small percentage of incompleteness ($\sim 15\%$) or contamination ($\sim 15\%$) while requiring a much narrower wavelength range.  
    
    \item Using an optimized combination of Lick indices (namely, \ind{4}, \ind{6}, \ind{7}, \ind{8}, \ind{9}, \ind{10}, \ind{11}, \ind{12}, \ind{14}, and \ind{15}), we measure single-burst stellar \age, \met, and \afe for 140 passive galaxies, without assuming cosmological priors on the maximum value of \age as a function of redshift. We also performed an extended analysis to assess the impact of different choices of indices, verifying that our findings are robust against the choice of a different combination of indices (\ref{app:combos}).
    
    \item We find trends between $\log\age$, \met, \afe, and the stellar velocity dispersion consistent with those expected from a passively evolving population, with slopes of ($0.5\pm0.1$), ($0.3\pm0.2$), and ($0.2\pm0.1$), respectively (Figure~\ref{fig:res:parsigma}). This analysis shows, for the first time using individual galaxies, that a relation between \afe\ (a star formation timescale proxy) and \sigmastar\ is already in place at $z\sim 0.7$. Moreover, the age difference of 5.5~Gyr between our sample and local ETGs can be entirely accounted for by a pure passive evolution. Assuming a standard $\Lambda$CDM cosmology, the relation between formation redshifts and galaxy stellar masses is found to agree with several previous analyses (Figure~\ref{fig:res:zf_mass}), confirming that this population of massive galaxies forms at $z_\mathrm{form}\sim 1.3(1.6)$ at masses $\log M_\star/M_\odot=10.7(11.3)$, after the peak of the cosmic star formation rate density. 
    
    \item Even if we do not impose any cosmological prior to the age of the population, the obtained \age--redshift evolution is consistent with a $\Lambda$CDM universe (Figure~\ref{fig:res:param_redshift}). The stellar \met\ and \afe\ do not evolve significantly over $z$ and their values are slightly supersolar, $\med{\met}=0.08\pm0.18\,(1\sigma)$~dex, and supersolar, $\med{\afe}=0.13\pm0.11\,(1\sigma)$~dex, compatibly with their local counterparts.

    \item Finally, the analysis of median binned relations confirms the downsizing scenario and the passive nature of this population. Remarkably, we obtain two clear nearly parallel $\age-z$ relations for both the higher ($\sigmastar\approx 230~\kms$) and the lower ($\sigmastar\approx 200~\kms$) mass regimes. This difference of $\Delta\age\approx 0.5$~Gyr can be interpreted as a delay in formation time between the two, with later formation epochs for the population of less massive galaxies. 

  \end{enumerate}

  Overall, our analysis of individual galaxies confirms the existence of a population of passively evolving galaxies at intermediate redshift that follows a downsizing pattern. In a subsequent paper, we will make use of their median $\age-z$ relation to derive constraints on cosmological parameters and obtain a direct measure of the Hubble parameter $H(z)$ using the cosmic chronometers approach. In this context, we also plan to study in more detail the effect of assuming different SFHs. In this way, we will be able to study for the first time at these redshifts the detailed stellar population properties of passive galaxies and their underlying cosmology, jointly. 
  
  In the more distant future, dedicated large spectroscopic surveys would be crucial to extend the present analysis, providing at the same time unique opportunities for gravitational wave astronomy and cosmology \citep{Palmese2019}.

\begin{acknowledgments}
  We thank the anonymous referee for the constructive comments and suggestions that helped in improving this paper.
  This work is based on data products from observations made with ESO Telescopes at the La Silla Paranal Observatory under program ID 194.AF2005(A-N). We thank the LEGA-C team for making their dataset public, Daniel Thomas for supplying higher-resolution SSP models than those available publicly, and Nicholas Scott for useful suggestions for the analysis of spectral indices. N.B. and M.M. acknowledge support from MIUR, PRIN 2017 (grant 20179ZF5KS). M.M., A.C., and L.P. acknowledge the grants ASI n.I/023/12/0 and ASI n.2018-23-HH.0. A.C. acknowledges the support from grant PRIN MIUR 2017 - 20173ML3WW\_001. 
\end{acknowledgments}

\software{\texttt{Astropy} \citep{AstropyCollaboration2018}; \texttt{ChainConsumer} \citep{Hinton2016}; \texttt{emcee} \citep{ForemanMackey2013}; \texttt{LOESS} \citep{Cappellari2013b}; \texttt{LtsFit} \citep{Cappellari2013a}; \texttt{Matplotlib} \citep{Hunter2007}; \texttt{Numpy} \citep{Harris2020}; \texttt{Scipy} \citep{Virtanen2020}; and \texttt{Topcat} \citep{Taylor2005}.}

\newpage
\bibliography{references}

\appendix
  \vspace*{-4.8em}
  \section{Measuring spectral indices with \texttt{PyLick}}\label{app:pylick}
    Absorption-line features produced by atomic and molecular absorbers in stellar photospheres can unlock crucial information about the physical properties of stellar populations. Here we present \texttt{PyLick}, a flexible tool to measure spectral indices and associated uncertainties in galaxy spectra that includes different index definitions and measuring methods. The code is entirely written in Python language and object oriented. We developed this tool in the view of upcoming observational facilities to facilitate the detailed analysis of high-resolution spectroscopy (e.g., VLT/MOONS, \citealt{Cirasuolo2020}; Subaru/PFS, \citealt{Tamura2016}), and a quicker analysis of lower-resolution spectroscopic surveys \citep[e.g., Euclid,][]{Laureijs2011}. 
    
    \texttt{PyLick} is currently based on five modules: ``io'' implements the built-in methods for spectra I/O and preliminary analysis;  ``indices'' loads the index library including passband definitions; ``measure'' contains the methods to derive indices values and errors; ``plot'' deals with data visualization. Ultimately, the module ``analysis'' contains two classes: \texttt{Galaxy}, which is optimized to analyze a single spectrum, and \texttt{Catalog}, optimized to perform the analysis of a bulk set of spectra.
    
    In the current version, 54 indices are already defined in the index library:
    \begin{itemize}\itemsep1pt
      \item UV line indices: 11 far-UV and 8 mid-UV (\citealt{Fanelli1992}, see also \citealt{Maraston2008});
      \item Line break indices B2640, B2900 \citep{Spinrad1997};
      \item The Mg$_\mathrm{UV}$ index \citep{Daddi2005};
      \item 4000~\AA\ discontinuity indices: D4000 and D$_{n}$4000 \citep[][respectively]{Bruzual1983,Balogh1999};
      \item Lick indices \citep{Worthey1997, Trager1998};
      \item CaII H and CaII K ``pseudo-Lick'' indices \citep{Fanfani2019};
      \item Generic indices CaT, PaT, and CaT$^\ast$ \citep{Cenarro2001}.
    \end{itemize}
    However, new indices can be easily introduced by defining the wavelength regions of interest and the measuring method in a custom library file. 
    
    Different measuring methods are included: \texttt{lick\_atomic} (Equation \ref{eq:ind_licks1}), \texttt{lick\_molecular} (Equation \ref{eq:ind_licks2}), \texttt{break\_nu} (Equation \ref{eq:ind_D4000}), and \texttt{break\_lb} (as Equation \ref{eq:ind_D4000}, but integrating over $F(\lambda)\mathrm{d}\lambda$, instead of $F(\nu)\mathrm{d}\nu$). More specific methods are directly defined in the ``measure'' module. This is the case for the $\rm{Mg_{UV}}$ index that traces the absorption bump present at 2640--2850~\AA\ \citep{Daddi2005}:
    \begin{equation}
      \rm{Mg_{UV}} = \frac{2\int^{2725}_{2625}F(\lambda)\rm{d}\lambda}{\int^{2625}_{2525}F(\lambda)\rm{d}\lambda+\int^{2825}_{2725}F(\lambda)\rm{d}\lambda},
    \end{equation}
    and for the calcium triplet index
    \begin{equation}
      \rm{CaT^\ast} = \rm{CaT} -0.93\, \rm{PaT},
    \end{equation}
    which traces the strength of the CaII lines ($\rm{CaT}$) corrected from the contamination by Paschen lines ($\rm{PaT}$), as presented in detail in \cite{Cenarro2001}.
    
    Index errors are evaluated following the signal-to-noise method proposed by \cite{Cardiel1998}. The code is also able to handle bad pixels, which should be passed as a Boolean array using the \texttt{spec\_mask} argument. The user can choose a bad-to-total pixel ratio (we considered \texttt{BPR}=0.15 in this work) above which the measurement is not performed. Otherwise, a zeroth- or first-order interpolation is done over the bad pixels prior to the measurement. Finally, publication-quality figures can be produced with the plotting routines (e.g., Figures~\ref{fig:sample:composite_350_passive} and \ref{fig:sample:HK_definition}).

    \texttt{PyLick} has been released as an open-source project and is available on GitLab with extensive documentation and notebook examples \footnote{The code is available at \url{https://gitlab.com/mmoresco/pylick/}}.

    \begin{figure*}[t]
      \centering
      \includegraphics[width=0.93\textwidth]{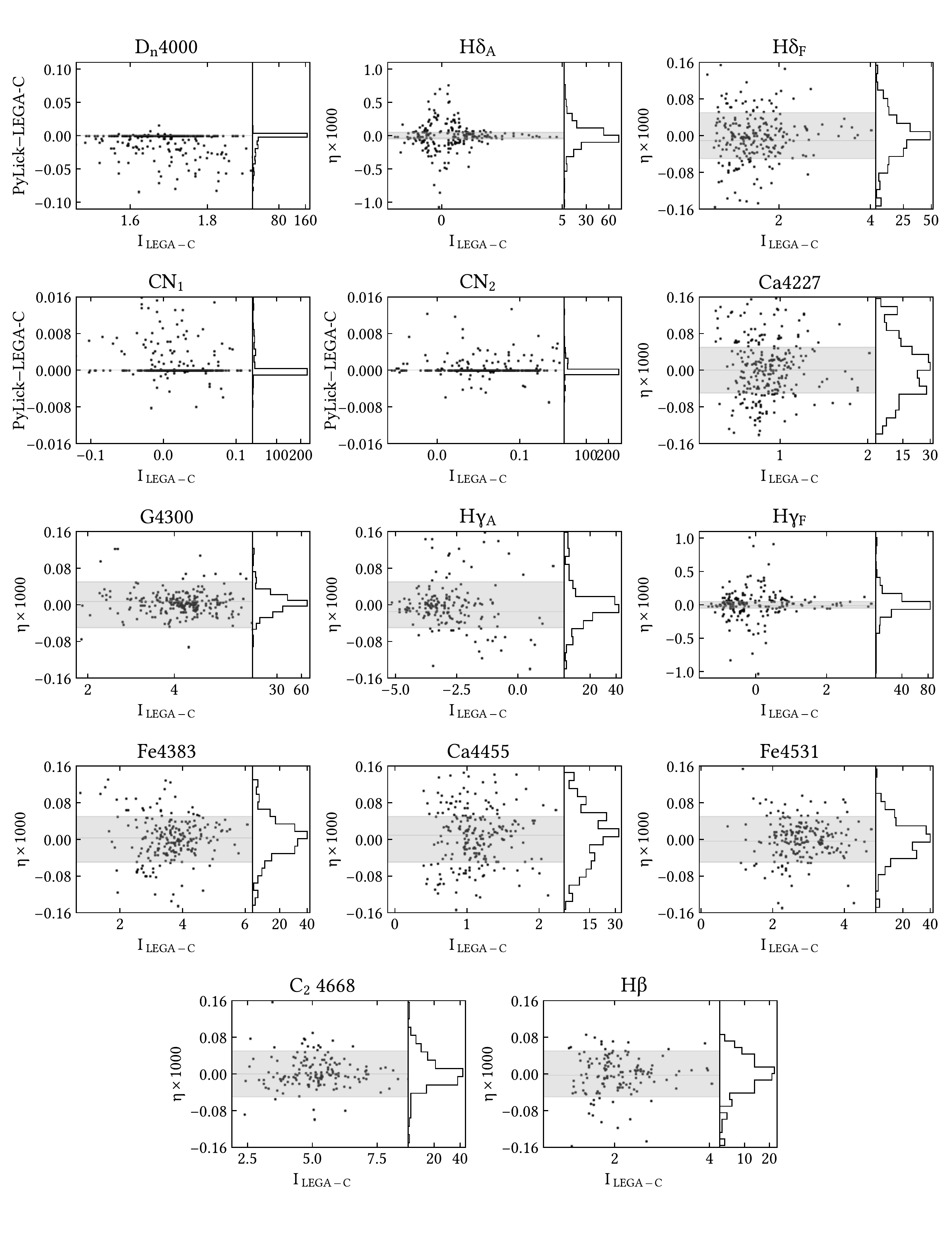}
      \vspace{-2.5em}
      \caption{Differences between the indices values measured with \texttt{PyLick} and those published in LEGA-C DR2 \citep{Straatman2018}, quantified as $\eta = (I_\mathrm{PyLick}-I_\mathrm{LEGA-C})/I_\mathrm{LEGA-C}$, except for the \ind{3} break and the molecular indices \ind{6} and \ind{7}, for which the absolute differences $(I_\mathrm{PyLick}-I_\mathrm{LEGA-C})$ are displayed. Gray bands correspond to the $\eta = \pm 5\times 10^{-4}$ regions. Note: $y$-axis limits are set to 5th--95th percentiles for illustrative purposes.\label{fig:app:pylick}}
    \end{figure*}

    \subsection{Validation of the code with LEGA-C data}
      To confirm the reliability and robustness of \texttt{PyLick}, we compare the measured indices and errors obtained on the unconvolved LEGA-C spectra $I_\mathrm{PyLick}$, with those released in the LEGA-C DR2 catalog $I_\mathrm{LEGA-C}$ \citep{Straatman2018}. The comparison is performed within the passive sample to minimize differences due to emission-line subtraction performed in LEGA-C DR2 \citep[see][]{Straatman2018}. Differences are computed as
      \begin{equation}
        \eta(I) = \frac{I_\mathrm{PyLick}-I_\mathrm{LEGA-C}}{\left|I_\mathrm{LEGA-C}\right|}.
      \end{equation}
      The same analysis is performed for uncertainties of the indices (the notation $\eta(\sigma)$ will be used). For a fair comparison, we multiplied our uncertainties for the same coefficients applied in the LEGA-C DR2 pipeline \citep[see][Tab.~3]{Straatman2018}. Figure~\ref{fig:app:pylick} shows the results for values of the indices, sorting each index available in LEGA-C DR2 by increasing wavelength.
      
      \begin{figure*}[t!]
        \centering
        \includegraphics[width=0.24\textwidth]{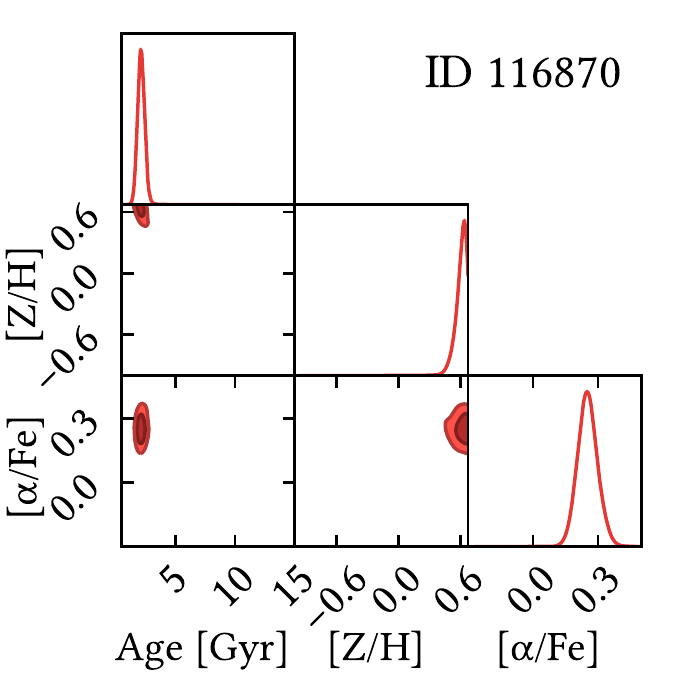} 
        \includegraphics[width=0.24\textwidth]{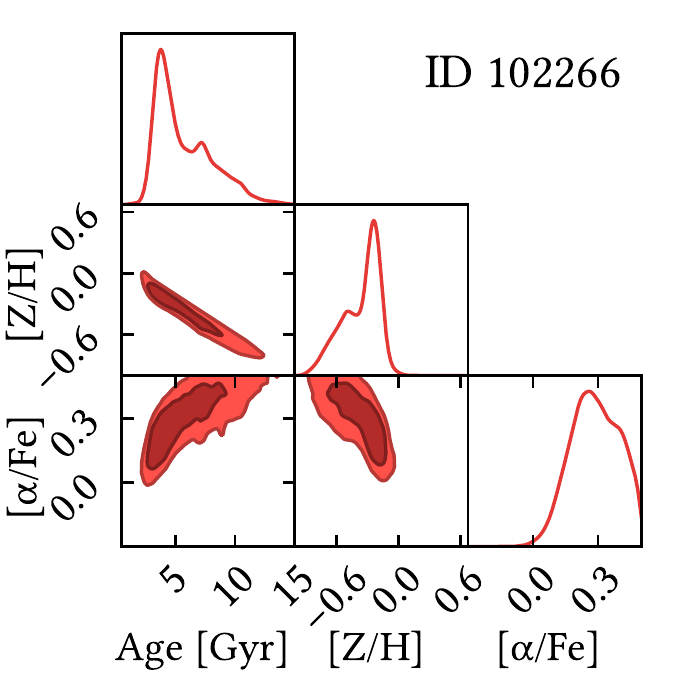}
        \includegraphics[width=0.24\textwidth]{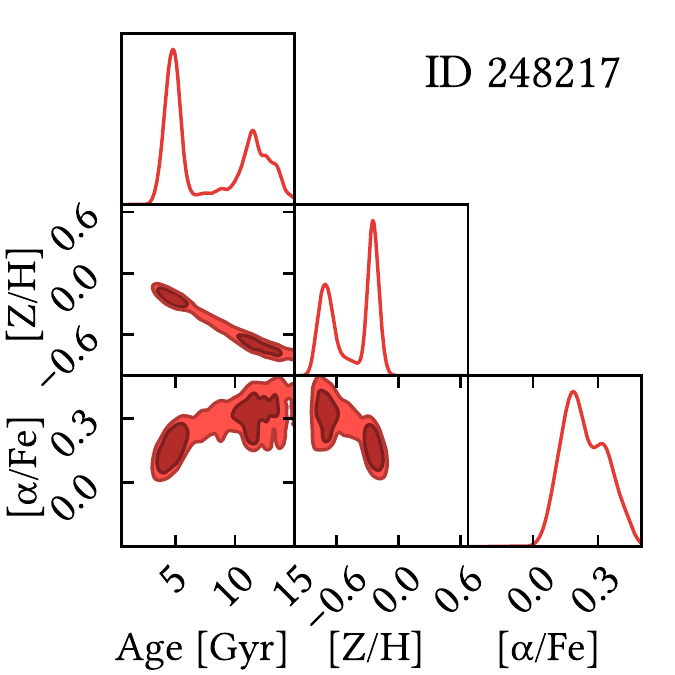}
        \includegraphics[width=0.24\textwidth]{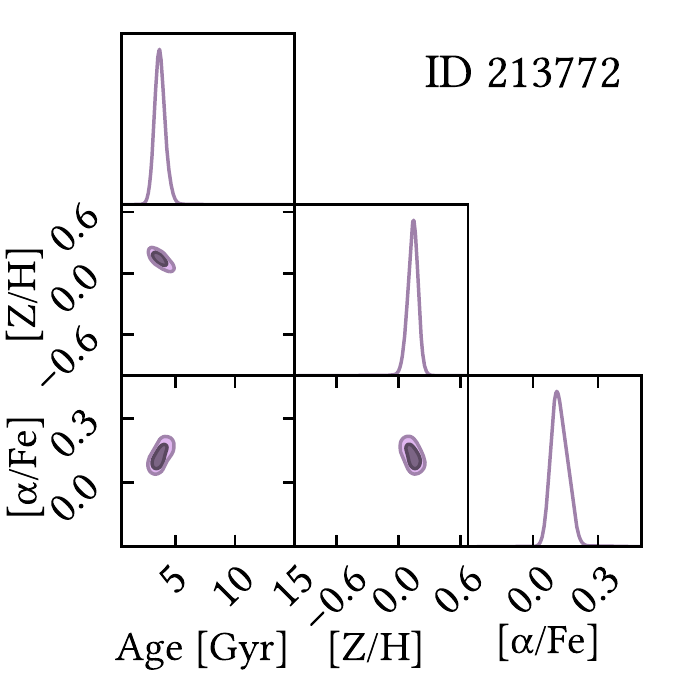}
        \caption{Examples of contour plots that may generate unreliable (red) and reliable (blue) results. Axes cover the full parameter space allowed from the models. Contours enclose 1 and 2$\sigma$ confidence regions.\label{fig:app_converged_examples}}
        \vspace*{0.5em}
      \end{figure*}

      Overall, we find excellent agreement with existing data, with a typical $\eta(I)$ of $\sim 5\times 10^{-5}$ ($\sim 10^{-4}$ $1\sigma$ scatter). Differences are lower for higher S/N indices \ind{9}, \ind{12}, \ind{14}, and \ind{15} ($|\eta(I)| < 10^{-5}$), and outliers are mostly galaxies with lower-quality spectra. There is no trend between relative differences and the indices values, and the distribution is qualitatively Gaussian. 
      
      We find good agreement also for the uncertainties, with a typical $\eta(\sigma)$ of $\sim 10^{-2}$  ($\sim 10^{-1}$ $1\sigma$ scatter). Larger deviations are seen for indices with values $\sim 0$ (Balmer indices, \ind{6}), for which measured uncertainties are $\sim 10\%$ lower. This difference can be due to different methods to estimate formal errors. 

      We note that at the current resolution ($R\sim 3500$), the method used to interpolate the spectra can introduce higher discrepancies than those observed before. In particular, using $0^{th}$ order interpolation the typical scatter in $\eta(I)$ increases up to $\sim 10^{-2}$. Overall, these results confirm the  reliability of \texttt{PyLick} to measure indices values and formal errors from observed spectra.

  \section{Assessing convergence and reliability of MCMC posterior distributions}\label{app:convergence}
    In a Bayesian analysis, it is crucial to determine whether MCMC chains are reproducing with sufficient accuracy the target posterior distribution. However, there is not an established standard to assess the convergence \citep{Hogg2018,Roy2020}. A possibility is to take into account the autocorrelation time $\tau_\mathrm{int}$ analysis, where $\tau_\mathrm{int}$ quantifies how many steps are needed to generate independent samples. We consider a chain to be formally converged when $\tau_\mathrm{int}$ for each parameter is greater than 1/100th of the chain size. In this work, analyzed galaxies typically require $\sim 7000$ steps. However, sample averages derived from formally converged chains could still be unreliable. This is the case when posterior distributions:
    \begin{enumerate}
      \item are skewed toward the priors (in this work, the limits of the parameter space allowed from the models);
      \item are not predictive (mainly due to high age-metallicity degeneracy);
      \item are multimodal.
    \end{enumerate}
    In Figure~\ref{fig:app_converged_examples} we show contour plots representative for the three categories and a typical contour plot for a good fit. We note that the age-metallicity degeneracy follows the so-called ``3/2 rule'' \citep{Worthey1998}, where an increase (decrease) of log age by a factor of 2, when accompanied by a decrease (increase) of stellar metallicity \met by a factor of 3, can reproduce the same set of observed indices. Multimodal distributions are obtained for less than 4\% of the analyzed galaxies and follow the direction of the age-metallicity degeneracy. As they do not recur for the same galaxy when slightly different sets of indices are considered, they should be considered as a special case of (2) and to be due to intrinsic degeneracies rather than a real complexity in the stellar population. We check joint and marginal distributions of all the 199 analyzed galaxies, and we flag and exclude galaxies belonging to these three categories. 

\begin{figure*}[t]
      \centering
      \includegraphics[width=0.99\textwidth]{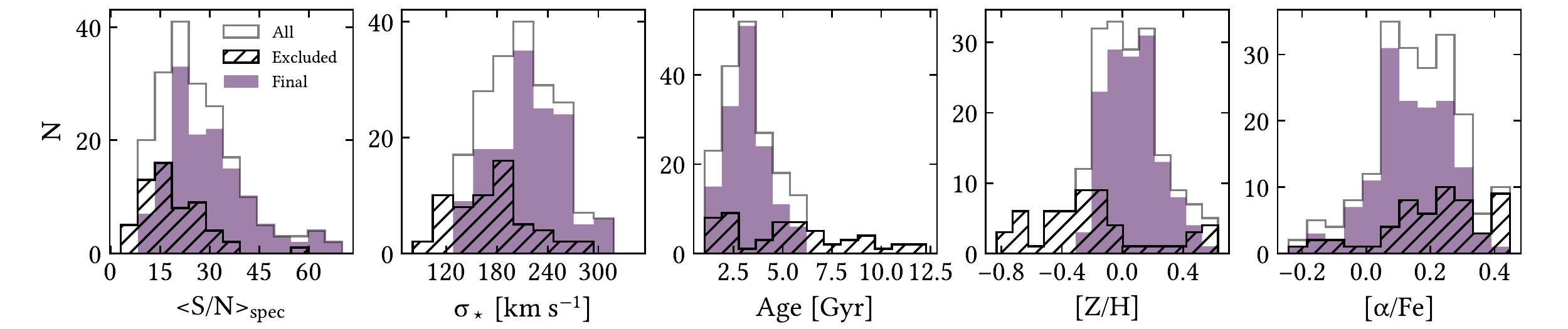}
      \includegraphics[width=0.99\textwidth]{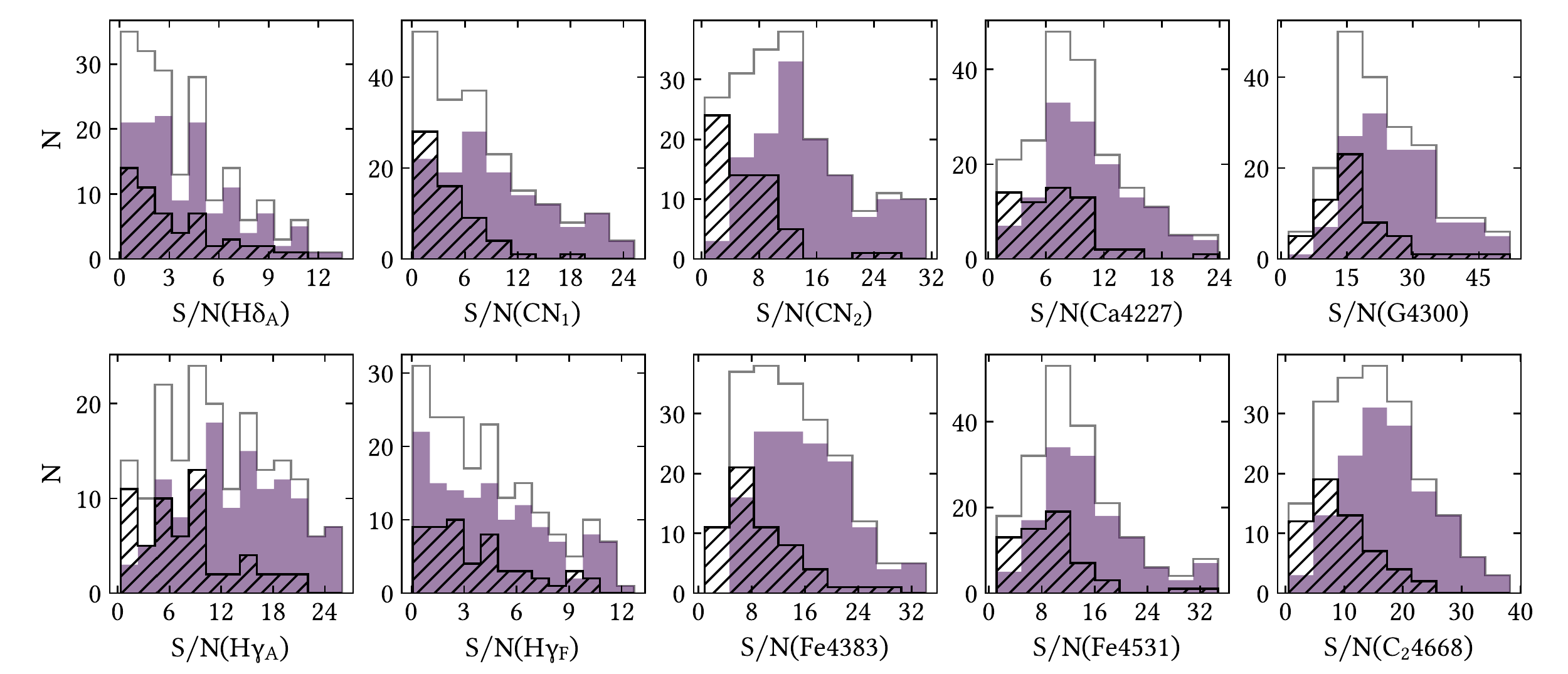}
      \caption{Distribution of overall spectral signal-to-noise ratio $\med{\rm S/N}_{spec}$, stellar velocity dispersion \sigmastar, stellar population parameters (\age, \met, \afe), and signal-to-noise ratios of individual indices for 199 analyzed passive galaxies in LEGA-C DR2 (gray histograms). Violet (black hatched) histograms represent galaxies included (excluded) after inspecting the MCMC posterior distributions.\label{fig:app_converged_distrib}}
    \end{figure*}

    In Figure~\ref{fig:app_converged_distrib} we show the distributions of the main parameters before and after this process. The great majority ($\sim 85\%$) of excluded galaxies have overall spectral $\med{\rm S/N}<25$. However, a cut of this kind applied \textit{a priori} would have halved the final sample, excluding also good converged fit. Low-$\sigmastar$ galaxies are preferentially excluded because they have relatively lower S/Ns. Distributions of stellar population parameters do not change significantly before and after this process. The posterior distribution of ages for galaxies with $\age >8$ Gyr tend to be heavily skewed toward the 15 Gyr prior and are also characterized by low metallicities ($\met < -0.4$) and relatively high $\alpha$/Fe values. 

    We observe that S/Ns of individual indices are a good indicator to separate included and excluded galaxies. For this purpose, we use indices with relatively higher S/N (see Table~\ref{tab:ind_prop}), namely, \ind{9}, \ind{12}, \ind{14}, and \ind{15}. The great majority ($\sim 90\%$) of excluded galaxies have at least one index with ${\rm S/N<10}$. A cut of this kind would still exclude $\sim 29\%$ of galaxies with reliable constraints. Anyway, we note that these cuts are not universal because they vary for different index combinations, and individual S/Ns of spectral indices vary for different spectral resolutions.

    In conclusion, we find that the inspection of MCMC posterior distributions is an efficient procedure to detect unreliable constraints--mostly due to low indices S/N--while maximizing the number of galaxies for which we obtain reliable constraints.

  \section{Different sets of indices}\label{app:combos}
    In this section we illustrate the results obtained with different index sets. We explore the huge number of viable combinations ($\sim 1$ million) following three approaches: 
    \begin{itemize}
      \item \textit{Maximize the information to be fitted}. Using a higher number of indices should provide more stable results. However, indices should be calibrated, and able to disentangle degeneracies giving equal weight to each model parameter. 
      \item \textit{Use already-proposed index combinations}. In particular we include redder Mg and Fe indices as done by previous works.
      \item \textit{Use a small, essential set to break the existing degeneracy between parameters}.
      Spectral indices are sensitive to variations in  $\age$, $\met$, $\afe$, but the relative sensitivity to these parameters is not identical. One can choose a small combination of four to six indices based upon their different sensitivity \citep[e.g., for element abundances see][]{Tripicco1995, Korn2005, Lee2009}.
    \end{itemize}

    As discussed in the text, we find an optimal combination that maximizes the spectral coverage, the number of constrained galaxies, and the precision of the constraints: \ind{4}, \ind{6}, \ind{7}, \ind{8}, \ind{9}, \ind{10}, \ind{11}, \ind{12}, \ind{14}, and \ind{15} (hereafter \textit{baseline}). In Table~\ref{table:app:index_sets}, we report the other main sets of indices that we analyzed, and the differences in the derived parameters with respect to the \textit{baseline} combination. We also analyzed many other sets ($\sim 50$), but they do not add significant information to this study. 

    \begin{splitdeluxetable*}{cccccccccccccccccccccBccccc}
      \tablewidth{0pt} 
      \tablenum{6}
      \tablecaption{Examples of Analyzed Index Combinations and Their Definitions.\label{table:app:index_sets}}
      \tablehead{\colhead{Combo ID} & \colhead{\rot{\ind{4}}}  & \colhead{\rot{\ind{5}}}  & \colhead{\rot{\ind{6}}} & \colhead{\rot{\ind{7}}}  & \colhead{\rot{\ind{8}}}  & \colhead{\rot{\ind{9}}}  & \colhead{\rot{\ind{10}}} & \colhead{\rot{\ind{11}}} & \colhead{\rot{\ind{12}}} & \colhead{\rot{\ind{13}}} & \colhead{\rot{\ind{14}}} & \colhead{\rot{\ind{15}}} & \colhead{\rot{\ind{16}}} & \colhead{\rot{\ind{17}}} & \colhead{\rot{\ind{18}}} & \colhead{\rot{\ind{19}}} & \colhead{\rot{\ind{20}}} & \colhead{\rot{\ind{21}}} & \colhead{\rot{\ind{22}}} & \colhead{\rot{\ind{23}}} & \colhead{Combo ID} & \colhead{N (in common)} & \colhead{$\Delta \age~(\sigma)$} & \colhead{$\Delta \met~(\sigma)$} & \colhead{$\Delta \afe~(\sigma)$}
      } 
      \startdata 
        \textit{baseline} & $\blacksquare$ &  & $\blacksquare$ & $\blacksquare$ & $\blacksquare$ & $\blacksquare$ & $\blacksquare$ & $\blacksquare$ & $\blacksquare$ &  & $\blacksquare$ & $\blacksquare$ &  &  &  & & &  &  &  & \textit{baseline} & 140 (140) & -- & -- & -- \\
        1 & $\blacksquare$ &  &  &  & $\blacksquare$ & $\blacksquare$ & $\blacksquare$ & $\blacksquare$ & $\blacksquare$ &  & $\blacksquare$ & $\blacksquare$ &  &  &  &  & & &  &  &  1 & 105 (95) & 0.04 (0.06) & -0.04 (0.42) & -0.05 (0.64) \\
        2 & $\blacksquare$ &  &  & $\blacksquare$ & $\blacksquare$ & $\blacksquare$ & $\blacksquare$ &  & $\blacksquare$ &  & $\blacksquare$ & $\blacksquare$ &  &  &  &  &  &  &  & &  2 & 131 (115) & -0.42 (0.63) & 0.03 (0.33) & -0.06 (0.76) \\
        3 & $\blacksquare$ &  & $\blacksquare$ & $\blacksquare$ & $\blacksquare$ & $\blacksquare$ & $\blacksquare$ & $\blacksquare$ & $\blacksquare$ &  & $\blacksquare$ & $\blacksquare$ &  &  &  &  & $\blacksquare$ &  &  & & 3 & 39 (39) & 0.05 (0.08) & 0.00 (0.02) & 0.08 (0.92) \\
        4 & $\blacksquare$ &  &  &  & $\blacksquare$ & $\blacksquare$ & $\blacksquare$ & $\blacksquare$ & $\blacksquare$ &  & $\blacksquare$ & $\blacksquare$ &  &  & $\blacksquare$ & $\blacksquare$ & $\blacksquare$ & $\blacksquare$ & $\blacksquare$ & $\blacksquare$ & 4 & 11 (8) & 0.25 (0.36) & -0.01 (0.09) & 0.09 (1.15)\\
        5 &  &  &  & $\blacksquare$ & $\blacksquare$ & $\blacksquare$ &  & $\blacksquare$ & $\blacksquare$ &  &  &  &  &  &  &  &  &  &  & & 5 & 133 (98) & -0.17 (0.25) & 0.01 (0.08) & -0.13 (1.61) \\
        6 &  &  &  & $\blacksquare$ & $\blacksquare$ & $\blacksquare$ &  &  & $\blacksquare$ & $\blacksquare$ &  & $\blacksquare$ &  &  &  &  &  & & & & 6 & 119 (102) & -1.11 (1.54) & 0.11 (0.95) & -0.06 (0.69) \\
      \enddata
      \tablecomments{For each combination we report the number of constrained galaxies, along with simple and $1\sigma$ differences in the derived parameters with respect to the \textit{baseline} combination.}
    \end{splitdeluxetable*}

    We find no significant systematic differences in the derived parameters when small changes to the baseline combination are applied, i.e. by adding or removing one to two indices. In particular, we focus here on the removal of CN indices, as the nitrogen abundance is not a free parameter in \citetalias{Thomas2011} models (Combo 1), and on the removal of those indices that sample twice the same spectral region (Combo 2). In the first case, we obtain constraints for fewer galaxies with respect to the \textit{baseline} combination, but with an overall excellent agreement. In the second case, we constrain about the same number of galaxies, obtaining lower ages but still in agreement with the \textit{baseline} set. This is likely due to the removal of \ind{11} that reduces the weight of age-sensitive features, therefore producing less reliable age estimates.

    \begin{figure}
      \centering
        \includegraphics[width=0.49\textwidth]{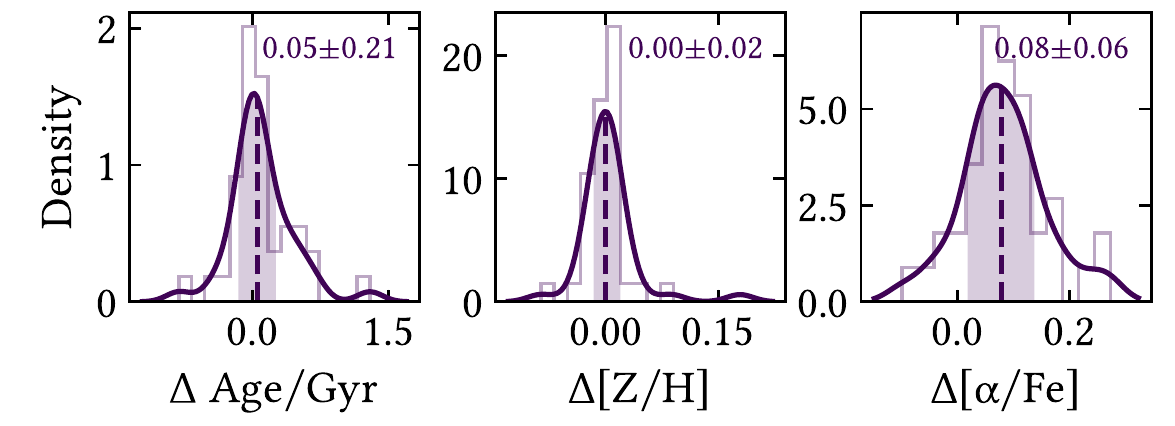}
      \caption{Differences between the parameters obtained with and without \ind{20} (Combo 3 vs. \textit{baseline}). Median (dashed line) and $\pm$NMAD values (shaded region) are annotated in the upper corners.\label{fig:analysis:cfr_combo_Mg}}
    \end{figure}

    Index combinations discussed above lack redder indices, such as the Mg indices, traditionally used as $\alpha$-abundance indicators. Therefore, we repeat the analysis including \ind{20} (Combo 3). Amongst the 59 with such relatively large spectral coverage, we obtain constraints for 39 galaxies. Differences are shown in greater detail in Figure \ref{fig:analysis:cfr_combo_Mg}. While ages and \met are in overall good agreement, the inclusion of another $\alpha$-sensitive index suggests that \afe values derived with Combo 1 may be underestimated by $\sim 0.1$ dex. However, given the small statistical significance, we do not correct for this offset. A similar discussion can be done by extending the analysis at all the redder indices (Combo 4, which is also the same combination used in \citealt{Onodera2015}).
      
    Finally, we note here that although minimal sets of $N=4$ indices in the wavelength region between 4000 and 4600~\AA\ allow us to analyze a large number of galaxies ($\sim300$), we do not find a relevant set to place constraints on more than one-third of them. This situation is improved when $N=5-7$ (e.g., Combo 5), but results show an overall stronger age–-metallicity degeneracy with respect to $N\geq 8$. It is also interesting to note that even if we do not include Balmer indices (Combo 6), we still obtain constraints for more than 100 galaxies. But we stress again that results are less reliable because we removed most of the age-sensitive features. 
          
    In summary, we find that a blind choice of index combinations can lead to less-robust results. This happens when a combination is unbalanced toward one or more parameters of the fit, but also if indices are measured on spectra where the sky subtraction was imperfect. After an extensive analysis, we demonstrate that within the limited statistics and wavelength coverage of current data, results do not show significant systematic differences.

\end{document}